\documentclass[preprint,3p,review]{elsarticle}

\usepackage{graphicx}
\usepackage{amssymb}
\usepackage{amsthm}
\usepackage{amsfonts}
\usepackage{amsbsy}
\usepackage{amsmath}
\usepackage{graphics}
\usepackage{csquotes}
\usepackage[labelformat=brace]{subcaption}
\usepackage[pdfencoding=auto, psdextra]{hyperref}
\hypersetup{colorlinks=true,linkcolor=blue, anchorcolor=red, citecolor=blue, filecolor = red, urlcolor = red,
            pdfauthor=author}
\usepackage{bookmark}
\usepackage[T1]{fontenc}
\usepackage{array, booktabs, makecell}
\usepackage{siunitx}
\usepackage{xurl}
\usepackage{multirow}
\usepackage{floatrow}
\usepackage{caption}
\usepackage{comment}

\journal{}
\newtheoremstyle{cited}%
  {3pt}
  {3pt}
  {\itshape}
  {}
  {\bfseries}
  {}
  {.5em}
  {\thmname{#1} \thmnumber{#2}\thmnote{ \normalfont#3}}
\theoremstyle{cited}

\newtheorem{theorem}{Theorem}[section]

\newtheorem{definition}{Definition}[section]

\newtheorem{Remark}{Remark}[section]

\begin{document}

\begin{frontmatter}


\title{On Multifractionality of Spherical Random Fields\\ with Cosmological Applications}



\author[b]{{Philip~Broadbridge}}
\ead{P.Broadbridge@latrobe.edu.au}

\author[b]{{Ravindi~Nanayakkara}}
\ead{D.Nanayakkara@latrobe.edu.au}

\author[b]{{Andriy~Olenko}\corref{cor1}}
\cortext[cor1]{Corresponding author}
\ead{A.Olenko@latrobe.edu.au}

\address[b]{Department of Mathematics and Statistics, La Trobe University, Melbourne, VIC 3086, Australia}

\begin{abstract}
This paper studies random fields on the unit sphere. Traditionally, isotropic Gaussian random fields are considered as the underlying statistical model of the cosmic microwave background (CMB) data. This paper discusses the generalized multifractional Brownian motion and its pointwise H\"older exponent on the sphere. The multifractional approach is used to investigate the CMB data from the Planck mission. These data consist of CMB radiation measurements at narrow angles of the sky sphere.  The obtained results suggest that the estimated H\"older exponents for different CMB regions do change from location to location. Therefore, CMB data are multifractional. Then the developed methodology is used to suggest two approaches for detecting regions with anomalies in cleaned CMB maps.

\
\end{abstract}


\end{frontmatter}

\section{Introduction}
\label{S2:1}

The notion of fractional Brownian motion (FBM) was introduced by B. Mandelbrot and Van Ness in~1968. The FBM depends on the Hurst parameter $H$, where $H \in (0, 1)$. The Hurst parameter can be used to define the H\"older regularity of FBM. The multifractional Brownian motion (MBM) was first considered by P\'eltier and L\'evy V\'ehel in 1995 extending the FBM, see~\cite{Ayache:2004}. The concept of multifractionality induces from fractionality allowing local properties to depend on space-time locations. The Hurst parameter $H$ of FBM is replaced by $H(t)$ in MBM. The MBM was proposed to model data that cannot be described by standard processes with stationary increments since their pointwise smoothness changes from point to point.

Multifractional processes were used to study complex stochastic systems which exhibit nonlinear behaviour in space and time. Multifractional behaviour of data has been found in many applications such~as, image processing, stock price movements, signal processing, see~\cite{Ayache:2004, Bianchi:2007, Sheng:2011}. Multifractional processes are more flexible in comparison to FBM and can be non-stationary. Multifractional Gaussian processes were studied in~\cite{Benassi:1998} where a method to evaluate the multifractionality using discrete observations of a process's single sample path was proposed. 

The generalized multifractional Brownian motion (GMBM) is a continuous Gaussian process that was introduced by generalizing the traditional FBM and MBM, see~\cite{Ayache:2004}. Comparing to MBM, the H\"older regularity of GMBM can substantially vary. For example, GMBM can allow discontinuous H\"older exponents. This has been an advantage to applications, specifically, medical image modelling, telecommunication, turbulence and finance where the pointwise H\"older exponent can change rapidly. A Fourier spectrum's low frequencies controls the long-range dependence of a stochastic process while the higher frequencies controls the H\"older regularity. Therefore, GMBM can be used to model processes that exhibit erratic behaviour of the local H\"older exponent and long-range dependence, see~\cite{Ayache:2004}. 

This paper deals with cosmological applications. The universe originated about 14 billion years ago and had extremely high temperature. The atoms were broken down into electrons and protons. The universe started to cool down and hydrogen atoms were formed 380,000 years after the Big Bang. As a result, photons were emitted and started moving without any restriction. This utmost ancient glow of light which is the leftover radiation after the Big Bang is called the cosmic microwave background radiation (CMB), see~\cite{Planck:2019}.  The CMB which dates back to nearly 400,000 years from the Big Bang was first discovered by Arno Pensiaz and Robert Wilson in 1964. The CMB is an electromagnetic radiation caused by the thermal movement of particles left in the universe. In the microwave region, the CMB spectrum closely follows that of a black body at equilibrium temperature 2.735K, tracing back to a plasma temperature  of around 4000K at a time corresponding to redshift  z=1500 at 50\% atomic combination. 

Although the equilibrium spectrum is important, there are important departures from equilibrium that give information on the state of the early universe, see for example~\cite{Pietroni:2009}. 
Relative anisotropic variations of spectral  intensity from that of a black body are of the order of~$10^{-4}$. Calculations by Khatri and Sunyaev~\cite{Khatri:2012} showed that outside of a relatively small range of redshifts, external energy inputs from sources such as massive particle decay, would dissipate by Compton and double Compton scattering and other relaxation processes to affect the signal by several lower orders of magnitude. The primary sources of anisotropy were large-scale acoustic waves whose compressions in the plasma universe were associated with raised temperatures. Using the current angular widths of anisotropies in the CMB, the current standard model $\Lambda$CDM (cold dark matter plus dark energy) affords an estimate of the Hubble constant at $H_0=67.4\pm 1.4$ km/s/MPc~\cite{Aghanim:2020}. This agrees well with data from the POLARBEAR Antarctica telescope that give $H_0=67.2\pm 0.57$ km/s/MPc~\cite{Adachi:2020}. However estimates from more recent emissions from closer galaxies using both cepheid variables and type Ia supervovae as distance markers, give $H_0=74.03\pm1.42$ km/s/MPc~\cite{Riess:2019}.  This unexplained discrepancy will eventually be resolved by newly found errors in the methodology of one or both of the competing large-z and small-z measurements, or in new physical processes that are currently unidentified.

Within a turbulent plasma, there are electrodynamical processes that are far more complicated than the large-scale acoustic waves. When radiation by plasma waves is taken into account, useful kinetic equations and spectral functions can no longer be constructed by Bogoliubov's approach of closing the moment equations for electron distribution functions, see Chapter~5 in~\cite{Klimontovich:1967}. Even in controlled tokomak devices, the dynamical description of magnetic field lines has fractal attracting sets~\cite{Viana:2011} and charged particle trajectories may have fractal attractors under the influence of multiple magnetic drift waves~\cite{Mathias:2017}. At CMB frequencies below 3 GHz (i.e. wavelengths larger than 10 cm), there have been indications of spectral intensities much higher than that of a 2.7 K black body~\cite{Baiesi:2020}. Although there is a high level of confidence in measuring the universe's expansion factor from CMB since the decoupling of photons from charged particles, the level of complexity of magnetohydrodynamics in plasma suggests that this subject might not be a closed book. Multi-fractal analysis is a tool that might contribute to understanding the multi-scale data that are becoming successively more fine-grained with each generation of radio telescope.

The space missions that have studied the CMB so far are Cosmic Background Explorer, Wilkinson Microwave Anisotropy Probe and Planck. The Planck mission was launched in 2009 to measure the CMB with an extraordinary accuracy over a wide spectrum of infrared wavelengths. The Planck mission traced the CMB anisotropies at narrow angles with a high resolution and sensitivity. The measured temperature intensities from the Planck mission together with the polarisation data can be used to check for the existence of anomalies within the CMB data. The CMB data can be utilized to understand how the early universe originated and to find out the key parameters of the Big Bang model, see~\cite{ESA:2019}. One of the key assumptions of the modern cosmology is that the universe looks the same in any direction. It has been in debate for several years by using the CMB data. Numerous research suggested that the CMB data are either non-Gaussian or cannot be accurately described by statistical or mathematical models with few constant parameters, see~\cite{Ade:2016, Hill:2018, Leonenko:2021, Marinucci:2004, Minkov:2019, Starck:2004}. The classical book by Weinberg~\cite{Weinberg:2008} explained that this anisotropy in the plasma universe was significant enough to produce anisotropy in current galaxy distributions. For some recent results and discussion of fundamental cosmological models of the universe, see~\cite{Deutscher:2020}. To detect departures from the isotropic model in actual CMB data several statistical approaches can be employed, see, for example,~\cite{Hamann:2021, Leonenko:2021}. Different approaches can result in different results and suggest to cosmologists sky regions for further investigations. The motivation of this paper is to check for multifractionality of the CMB data from the Planck mission. 

Theoretical multifractional space-time models which differ from the standard cosmological model have been studied by~\cite{Calcagni:2019,Calcagni:2016}. They proposed several theoretical advancements using multifractional space-time that change its properties from place to place. It was suggested that the universe is not expanding monotonically which produces multifractional behaviour. In~\cite{Calcagni:2016}, the CMB data from Planck mission and Far Infrared Absolute Spectrophotometer were used to establish speculative constraints on multifractional space-time expansion scenarios. Further, fractional SPDEs were employed to model the CMB data from Planck mission and study their changes, see~\cite{Anh:2018}. The considered fractional SPDE models exhibited long-range dependence.

This paper uses the theory of multifractional random fields  and  develops methodology to investigate fractional properties of random fields on the unit sphere. The presented detailed  analysis of actual CMB data suggests the presence of multifractionality. 

The developed methodology was also used to detect anomalies in CMB maps. The obtained results were compared with a different method from~\cite{Hamann:2021}. It was demonstrated that the both methods can find same anomalies, but each method also can detect its own CMB regions of unusual behaviour. It was shown that applications of the developed methodology resulted in spatial clusters with high values of the proposed discrepancy statistics. The clusters matched very well with the TMASK of unreliable CMB intensities.

The structure of the paper is as follows. Section~\ref{S2:2} provides main notations and definitions related to the theory of random fields. Section~\ref{S2:3} introduces the concept of multifractionality and discusses the GMBM. Section~\ref{S2:4} presents results on the estimation of the pointwise H\"older exponent by using quadratic variations of random fields. Section~\ref{S2:5} discusses the suggested estimation methodology. Numerical studies including computing the estimates of pointwise H\"older exponents for different one- and two-dimensional regions of the CMB sky sphere are given in Section~\ref{S2:6}. This section also demonstrates an application of the developed methodology to detect regions with anomalies in the cleaned~CMB~maps. Finally, the conclusions and some future research directions are presented in~Section~\ref{S2:7}. 

All numerical studies were carried out by using the software Python version 3.9.4 and R version 4.0.3, specifically, the R package~{\sc rcosmo} \cite{Fryer:2020,rcosmo:2019}. A reproducible version of the code in this paper is available in the folder \enquote{Research materials} from the website~\url{https://sites.google.com/site/olenkoandriy/}.

\section{Main notations and definitions}
\label{S2:2}

This section presents background material in the theory of random fields, fractional spherical fields and fractional processes. Most of the material included in this section is based on \cite{Ayache:2018, GARCIA:2020, Herbin:2006, Lang:2015, Malyarenko:2012, Marinucci:2011}.

Let $\mathbb{R}^{3}$ be the real 3-dimensional Euclidean space and $s_2(1)$ be the unit sphere defined in $\mathbb{R}^{3}$. That is, $s_2(1)=\left\{x \in \mathbb{R}^{3},\|x\|=1\right\}$ where $\|\cdot\|$ represents the Euclidean distance in ${\mathbb{R}}^3$. Let ${SO}(3)$ denotes the group of rotations on $\mathbb{R}^{3}$.

Let $(\Omega, \mathcal{A}, P)$ be a probability space. The symbol $\overset{d}{=}$  denotes the equality in the sense of the finite-dimensional~distributions.

\begin{definition}
A function $T(\omega, x): \Omega \times s_2(1) \rightarrow \mathbb{R}$ is called a real-valued random field defined on the unit sphere. For simplicity, it will also be denoted by $T(x)$, $x \in s_2(1)$.
\end{definition}

\begin{definition}
The random field $T(x)$ is called strongly isotropic if for all $k \in \mathbb{N}$, $x_{1}, \ldots, x_{k} \in s_2(1)$ and $g \in {SO}(3)$, the joint distributions of the random variables $T\left(x_{1}\right), \ldots, T\left(x_{k}\right)$ and $T\left(g x_{1}\right), \ldots, T\left(g x_{k}\right)$ have the same law.

It is called $2$-weakly isotropic (in the following, it will be just called  isotropic) if the second moment of $T(x)$ is finite, i.e. if $E\Big(|T(x)|^{2}\Big)<+\infty$ for all $x \in s_2(1)$ and if for all pairs of points $x_{1}, x_{2} \in s_2(1),$ and for any rotation, $g \in {SO}(3)$, it holds
\[
E\Big(T(x)\Big) = E\Big(T(gx)\Big),\quad 
E\Big(T\left(x_{1}\right) T\left(x_{2}\right)\Big) = E\Big(T\left(gx_{1}\right) T\left(gx_{2}\right)\Big).
\]
\end{definition}

\begin{definition}
$T(x)$ is called Gaussian if for all $k \in \mathbb{N}$ and $x_{1}, \ldots, x_{k} \in s_2(1)$ the random variables $T\left(x_{1}\right), \ldots, T\left(x_{k}\right)$ are multivariate Gaussian distributed; that is, $\sum_{i=1}^{k} a_{i} T\left(x_{i}\right)$ is a normally distributed random variable for all $a_{i} \in \mathbb{R}$, $i=1, \ldots, k,$ such that $\sum_{i=1}^{k} a_i^2 \neq 0.$
\end{definition}

Let $T = \{ T(r,\theta,\varphi) : 0 \leq \theta \leq \pi, 0 \leq \varphi \leq 2\pi, r > 0\}$ be a spherical random field that has zero mean, finite variance and is mean square continuous. Let the corresponding Lebesgue measure on the unit sphere be $\sigma_1(du)
    = \sigma_1(d\theta \cdot d\varphi) = \sin{\theta}d\theta d\varphi$, $u = (\theta,\varphi) \in s_2(1)$. For two points on $s_2(1)$, we use  $\Theta$ to denote the angle formed between two rays originating at the origin and pointing at these two points. $\Theta$ is called the angular distance between these two points. To emphasize that a random field depends on Euclidean coordinates, the notation $\Tilde{T}(x) = T(r,\theta,\varphi)$, $x \in \mathbb{R}^3$, will be used.
    
\begin{Remark}
A real-valued second order random field $\Tilde{T}(x)$, $x \in s_2(1),$ with $E\Big(\Tilde{T}(x)\Big)=0$ is isotropic if $E\Big(\Tilde{T}(x_1)\Tilde{T}(x_2)\Big) = B(\cos{\Theta})$, $x_1, x_2 \in s_2(1)$, depends only on the angular distance $\Theta$ between $x_1$ and $x_2$.
\end{Remark}

The spherical harmonics are defined by
\[Y_l^m (\theta,\varphi) = c_l^m\exp{(im \varphi)}P_l^m(\cos{\theta}), \quad l=0,1,..., \; m=0, \pm 1, ..., \pm l,\]
with
$$c_l^m = (-1)^m \left(\frac{2l+1}{4\pi}\frac{(l-m)!}{(l+m)!}\right)^{1/2},$$ and the Legendre polynomials $P_l^m(\cos{\theta})$ having degree $l$ and order $m$.

Then the following spectral representation of spherical random fields holds in the mean-square sense:
\[
T(r, \theta, \varphi)=\sum_{l=0}^{\infty} \sum_{m=-l}^{l} Y_{l}^{m}(\theta, \varphi) a_{l}^{m}(r),
\]
where $a_{l}^{m}(r)$ is a set of random coefficients defined by
\[
    a_l^m(r)=\int_0^{\pi} \int_0^{2\pi} T(r,\theta,\varphi)\overline{Y_l^m(\theta, \varphi)}r^2 \sin{\theta}d\theta d\varphi = \int_{s_2(1)}\Tilde{T}(ru)\overline{Y_l^m(u)}\sigma_1(du), 
\]
where $u= {\frac{x}{\Vert x \Vert}} \in s_2(1)$, $r= \Vert x \Vert$. 

\begin{definition}
A real-valued random field $\Tilde{T}(x)$, $x \in \mathbb{R}^{3}$, is with stationary increments if the equality
\[
\Tilde{T}(x+{x^{\prime}})-\Tilde{T}({x^{\prime}}) \stackrel{d}{=} \Tilde{T}(x)-\Tilde{T}(0), \; x \in \mathbb{R}^{3},
\]
holds, for all ${x^{\prime}} \in \mathbb{R}^{3}$.
\end{definition}

\begin{Remark}
   When $\Tilde{T}(x)$, $x \in \mathbb{R}^{3}$, is a second order random field with stationary increments, then one has,
\[
E\Big(\Tilde{T}(x+x^{\prime})-\Tilde{T}(x^{\prime})\Big)^{2}=\mathcal{V}_{\Tilde{T}}(x), \quad \text{ for every }(x, x^{\prime}) \in \mathbb{R}^{3} \times \mathbb{R}^{3},
\]
where $\mathcal{V}_{\Tilde{T}}$ is called the variogram of the field $\Tilde{T}$.
\end{Remark}

\begin{definition}
A real-valued random field $\Tilde{T}(x)$, $x \in \mathbb{R}^{3}$, is said to be globally self-similar, if for some fixed positive real number $H$ and for each
positive real number $a$, it satisfies
\begin{equation}
    a^{-H} \Tilde{T}(a x) \stackrel{d}{=} \Tilde{T}(x), \;  x \in \mathbb{R}^{3}.\label{eq:1b}
\end{equation}
\end{definition}

\begin{Remark}
Beside the degenerate case, the scale invariance property \eqref{eq:1b} holds only for a unique $H$ which we declare as the global self-similarity exponent.
\end{Remark}

\begin{definition}[\cite{Ayache:2018}]
For each fixed $H \in(0,1),$ there exists a real-valued globally $H$-self-similar isotropic centered Gaussian field with stationary increments. Up to a multiplicative constant, this field is unique in distribution. It is called fractional Brownian field (FBF) of Hurst parameter $H,$ and denoted by $B_{H}(t)$, $t \in \mathbb{R}^{3}$. The corresponding covariance function, is given, for all $\left(t^{\prime}, t^{\prime \prime}\right) \in \mathbb{R}^{3} \times \mathbb{R}^{3}$, by
\[
E\Big(B_{H}\left(t^{\prime}\right) B_{H}\left(t^{\prime \prime}\right)\Big)=2^{-1} \operatorname{Var}\left(B_{H}\left(e_{0}\right)\right)\left(\left\| t^{\prime}\right\|^{2 H}+\left\|t^{\prime \prime}\right\|^{2 H}-\left\|t^{\prime}-t^{\prime \prime}\right\|^{2 H}\right),
\]
where $e_{0}$ denotes an arbitrary vector of the unit sphere $s_2(1)$.
\end{definition}

\begin{Remark}
   In the particular case where $H=1 / 2,$ FBF is denoted by $B(t)$, $t \in \mathbb{R}^{3}$, and called L\'evy Brownian Motion. 
\end{Remark}

Similarly, one can introduce a $H\text{-self-similar}$ process in the one-dimensional case. We also denote it by $B_{H}(t)$, $t \geq 0$. It will be called the fractional Brownian motion (FBM).

\begin{definition}[\cite{Peltier:1995}]
The FBM with Hurst index $H(0<H<1)$ is defined as the stochastic integral
\[
B_{H}(t) = \frac{1}{\Gamma(H+1 / 2)}\left\{\int_{-\infty}^{0}\Big((t-s)^{H-1 / 2}-(-s)^{H-1 / 2}\Big) \mathrm{d} W(s) +\int_{0}^{t}(t-s)^{H-1 / 2} \mathrm{d} W(s)\right\}, \quad t \geq 0,
\]
where $W(\cdot)$ denotes a Wiener process on $(-\infty, \infty)$.
\end{definition}

The Hurst index $H$ is also known as the Hurst parameter which specifies the degree of self-similarity. When $H=0.5$, FBM reduces to the standard Brownian motion. In contrast to the Brownian motion, the increments of FBM are correlated. The FBM process $B_{H}(t)$ has the covariance function
\[
Cov\Big(B_{H}(s), B_{H}(t)\Big)=\frac{\sigma^{2}}{2}\left(|t|^{2H}+|s|^{2H}-|t-s|^{2H}\right).
\]

The mean value of FBM is ${E}\Big(B_{H}(t)\Big)=0$ and the variance function of FBM is ${Var}\Big(B_{H}(t)\Big)={\sigma^{2}|t|^{2H}}/2.$
The FBM has the following properties,
\begin{enumerate}[(i)]
    \item Stationary increments:
$
B_{H}(t)-B_{H}(s) \stackrel{d}{=}  B_{H}(t-s).$

    \item Long-range dependence of increments:
$
\sum_{n=1}^{\infty} E\Big({B_{H}(1)}(B_{H}(n+1)-B_{H}(n))\Big)=\infty, \quad H>1 / 2.
$
    \item Self-similarity:
$
B_{H}(a t) \stackrel{d}{=} |a|^{H} B_{H}(t).
$
\end{enumerate}

\section{Multifractional processes}
\label{S2:3}

This section provides definitions and theorems related to multifractional processes. Most of the material presented in this section is based on \cite{Ayache:2013, Ayache:2018, Benassi:1998, Benassi:1997, Peltier:1995}.

Let $C^1$ be the class of continuously differentiable functions and $C^2$ be the class of functions where both first and second derivatives exist and are continuous. 

First, we introduce multifractional processes in the one-dimensional case. They will be used to analyse CMB data using the ring ordering HEALPix scheme.

\begin{definition}[\cite{Benassi:1998}]
Multifractional Gaussian processes (MGP) $X(t), \; t \in [0,1],$ are real Gaussian processes whose covariance function $C(t,s)$ is of the form
\[
C(t, s)=\int_{\mathbb{R}} f(t, \lambda) \overline{f(s, \lambda)} \mathrm{d} \lambda,
\]
where
\[
f(t, \lambda)=\frac{\left(\mathrm{e}^{i t \lambda}-1\right) a(t, \lambda)}{|\lambda|^{1 / 2+\alpha(t)}}.
\]
\end{definition}
The smoothness of the process is determined by the function $\alpha(\cdot)$ which is from $C^{1}$ with $0<\alpha(t)<1$, $t \in[0,1]$. The modulation of the process is determined by the function $a(t, \lambda)$ which is defined on $[0,1] \times \mathbb{R}$ and satisfies
$
a(t, \lambda)=a_{\infty}(t)+R(t, \lambda),
$
where $a_{\infty}(\cdot)$ is $C^1([0,1])$ with, $a_{\infty}(t) \neq 0$ for all $t \in[0,1],$ and $R(\cdot, \cdot) \in C^{1,2}([0,1] \times \mathbb{R})$ is such that there exists some $\eta>0$ that for $i=0,1$ and $j=0,2$ it holds
\[
\left|\frac{\partial^{i+j}}{\partial t^{i} \partial \lambda^{j}} R(t, \lambda)\right| \leqslant \frac{C}{|\lambda|^{\eta+j}}.
\]

\begin{definition}[\cite{Peltier:1995}]
The multifractional Brownian motion (MBM) is given by
\[
B_{H(t)}(t)=\frac{\sigma}{\Gamma(H(t)+1 / 2)}\left\{\int_{-\infty}^{0}\Big((t-s)^{H(t)-1 / 2}-(-s)^{H(t)-1 / 2}\Big) \mathrm{d} B(s)
+\int_{0}^{t}(t-s)^{H(t)-1 / 2} \mathrm{d} B(s)\right\},
\]
where $B(s)$ is the standard Brownian motion and $\sigma^{2}= Var \Big(B_{H(t)}(t)\Big)|_{t=1}$.
\end{definition} 

For the MBM, ${E}\Big(B_{H(t)}(t)\Big)=0$ and 
${Var}\Big({B}_{{H}({t})}(t)\Big)={\sigma^{2}|{t}|^{2 {H}({t})}}/2$.

The FBM is a special case of the MBM where the local H\"older exponent $H(t)$ is a constant, namely, $H(t)=H$. The MBM which is a non-stationary Gaussian process does not have independent stationary increments in contrast to the FBM. 

\begin{definition}
A function $H(\cdot): \mathbb{R} \rightarrow \mathbb{R}$ is a $(\beta, c)$-H\"older function, $\beta>0$ and $c>0,$ if
\[
\left|H\left(t_{1}\right)-H\left(t_{2}\right)\right| \leqslant c\left|t_{1}-t_{2}\right|^{\beta},
\]
for all $t_{1}, t_{2}$ satisfying $\left|t_{1}-t_{2}\right|<1$.
\end{definition}

The MBM admits the following harmonizable representation, see~\cite{Benassi:1997}. If $H(\cdot): \mathbb{R} \rightarrow[a, b] \subset(0,1)$ is a $\beta$-H{\"o}lder function satisfying the assumption $\sup H(t)<\beta$, then the MBM with functional parameter $H(\cdot)$ can be written as $\operatorname{Re}\left(\int_{\mathbb{R}} \frac{\left(\mathrm{e}^{it \xi}-1\right)}{\|\xi\|^{H(t)+ 1 / 2}} \mathrm{d} \tilde{W}(\xi)\right),$ where ${\tilde{W}}(\cdot)$ is the complex isotropic random measure that satisfies $\mathrm{d}{\tilde{W}}(\cdot)=\mathrm{d} {W_{1}}(\cdot)+\mathrm{id}{W_{2}}(\cdot)$. Here, ${W_{1}}(\cdot)$ and ${W_{2}}(\cdot)$ are independent real-valued~Brownian measures.

Now we introduce the concept of the generalized multifractional Brownian motion (GMBM). The GMBM is an extension of the FBM and MBM. The GMBM was introduced to overcome the limitations existed in applying the MBM to model data whose pointwise H\"older exponent has an irregular~behaviour.

The following definitions will be used to analyse the CMB data using the ring and nested ordering HEALPix schemes for $d=1, 2$ respectively.

\begin{definition}[\cite{Ayache:2004}]
Let $[a, b] \subset(0,1)$ be an arbitrary fixed interval. $A n$ admissible sequence $\left(H_{n}(\cdot)\right)_{n \in \mathbb{N}}$ is $a$
sequence of Lipschitz functions defined on $[0,1]$ and taking values in $[a, b]$ with Lipschitz constants $\delta_{n}$ such
that $\delta_{n} \leqslant c_{1} 2^{n \alpha}$, for all $n \in \mathbb{N}$, where $c_{1}>0$ and $\alpha \in(0, a)$ are constants.
\end{definition}

\begin{definition}[\cite{Ayache:2004}]\label{def:42}
Let $\left(H_{n}(\cdot)\right)_{n \in \mathbb{N}}$ be an admissible sequence. The generalized multifractional field (GMF) with the parameter sequence $\left(H_{n}(\cdot)\right)_{n \in \mathbb{N}}$ is the continuous Gaussian field $Y(x, y), \; (x, y) \in [0,1]^{d} \times[0,1]^{d}$ defined for all $(x, y)$ as
\[
    Y(x, y)=\operatorname{Re}\left(\int_{\mathbb{R}^{d}}\left(\sum_{n=0}^{\infty} \frac{\left(\mathrm{c}^{ix\xi}-1\right)}{\|\xi\|^{H_{n}(y)+ 1/2}} \hat{f}_{n-1}(\xi)\right) \mathrm{d} \tilde{W}(\xi)\right),
\]
where ${\tilde{W}}(\cdot)$ is the stochastic measure defined previously.

The GMBM with the parameter sequence $\left(H_{n}(\cdot)\right)_{n \in \mathbb{N}}$ is the continuous
Gaussian process $X(t), \; t \in [0,1]^{d}$ defined as the restriction of $Y(x, y)$, $(x, y) \in {[0,1]}^{d} \times {[0,1]}^{d}$ to the diagonal: $X(t)=Y(t, t)$.
\end{definition}

Compared to the FBM and MBM, one of the major advantages of the GMBM is that its pointwise H{\"o}lder exponent can be defined through the parameter $\left(H_{n}(\cdot)\right)_{n \in \mathbb{N}}$. For every $t \in \mathbb{R}^{2},$ almost~surely,
\[
\alpha_{X}(t)=H(t)=\liminf _{n \rightarrow \infty} H_{n}(t).
\]

\section{The H{\"o}lder exponent}
\label{S2:4}

This section presents basic notations, definitions and theorems associated with the pointwise H{\"o}lder exponent, see \cite{Ayache:2004, Benassi:1998, Istas:1997} for additional details. The pointwise H{\"o}lder exponent determines the regularity of a stochastic process. It describes local scaling properties of random fields and can be used to detect multifractionality.

\begin{definition}\label{def2.4.1}
The pointwise H{\"o}lder exponent of a stochastic process ${X(t)}$, $t \in \mathbb{R},$ whose trajectories are continuous, is the stochastic process ${\alpha_{X}(t)}$, ${t \in \mathbb{R}}$, defined for every $t$ as
\[
\alpha_{X}(t)=\sup \left\{\alpha: \limsup _{h \rightarrow 0} \frac{|X(t+h)-X(t)|}{|h|^{\alpha}}=0\right\}.
\]
\end{definition}

The H{\"o}lder regularity of FBM can be specified at any given point $t$, almost surely and $\alpha_{B_H}(t)=H$ is constant for FBM. The pointwise H{\"o}lder regularity of MBM can be determined by its functional parameter similar to FBM where $\alpha_{Z}(t)$ is the pointwise H\"older exponent. Particularly, for every $t \in \mathbb{R}$, almost surely, $\alpha_{Z}(t)=H(t)$.

In literature, the method of quadratic variations is a frequently used technique to estimate the H{\"o}lder exponent, see~\cite{Benassi:1998, Istas:1997}. The following definition is used to compute the total increment in the one-dimensional case and will be applied for the ring ordering scheme of HEALPix points.

\begin{definition}\label{def2.4.2}
Let $t \in [0,1]$. For every integer $N \geq 2$,  the generalized quadratic variation $V_{N}^{(1)}(t)$ around $t$ is defined by
\begin{equation}
    V_{N}^{(1)}(t) = \sum_{p \in v_{N}(t)}\left(\sum_{k \in F}e_k X\left(\frac{p+k}{N}\right)\right)^2,\label{eq:2b}
\end{equation}
where $F=\{0,1,2\}$, $e_{0}=1$, $e_{1}=-2$ and $e_{2}=1$ and $v_{N}(t)=\left\{p \in \mathbb{N}: 0 \leqslant p \leqslant N-2 \text { and }\left|t -{p}/{N}\right| \leqslant N^{-\gamma}\right\}$.
\end{definition}

The following definition is used to compute the total increment in the two-dimensional case and will be used for the nested ordering scheme of HEALPix points.

\begin{definition}\label{def2.4.3}
Let $t = (t_1,t_2)\in [0,1]^2$. For every integer $N \geq 2$,  the generalized quadratic variation $V_{N}^{(2)}(t)$ around $t$ is defined by
\begin{equation}
    V_{N}^{(2)}(t) = \sum_{p \in v_{N}(t)}\left(\sum_{k \in F}d_k X\left(\frac{p+k}{N}\right)\right)^2,\label{eq:3b}
\end{equation}
where $p={\left(p_1,p_2\right)}$, $\varepsilon={\left(\varepsilon_1,\varepsilon_2\right)}$ and ${(p+\varepsilon)}/{N}=\left({{(p_1+\varepsilon_1)}/{N}},{{(p_2+\varepsilon_2)}/{N}}\right)$, $F=\{0,1,2\}^{2}$ and for all $k=\left(k_{1}, k_{2}\right) \in F$, $d_{k}=\prod_{l=1}^{2} e_{k_{l}}$ with $e_{0}=1$, $e_{1}=-2$ and $e_{2}=1$. Here,
$
v_{N}(t)=v_{N}^{1}\left(t_{1}\right) \times v_{N}^{2}\left(t_{2}\right)$
and for all $i=1, 2$, $
v_{N}^{i}\left(t_{i}\right)=\left\{p_{i} \in \mathbb{N}: 0 \leqslant p_{i} \leqslant N-2 \text { and }\left|t_{i}-{p_{i}}/{N}\right| \leqslant N^{-\gamma}\right\}
$.
\end{definition}

The pointwise H{\"o}lder exponents are estimated for the one-dimensional ring ordering and two-dimensional nested ordering of HEALPix points by considering sufficiently large $N$ and $d=1,2$ respectively in the following Theorem which is a specification of Theorem 2.2 in~\cite{Ayache:2004} with $\delta=1$.

\begin{theorem}[\cite{Ayache:2004}]\label{theo2.4.1}
Let $X(t), \; t \in [0,1]^d$, be a GMBM with an admissible sequence $(H_n(\cdot))_{n \in \mathbb{N}}$ ranging in $[a, b] \subset (0,1 - 1/2d).$ Then, for a fixed $\gamma \in (b, 1-1/2d)$ and the sequence $(H_n(t))_{n \in \mathbb{N}}$ convergent to $H(t)$, it almost surely holds
\begin{equation}
    H(t) = \lim_{N \to \infty} \frac{1}{2}\left(d(1-\gamma)-\frac{\mathrm{log} \Big(V_{N}^{(d)}(t)\Big)}{\mathrm{log}(N)}\right).\label{eq:4b}
\end{equation}
\end{theorem}

\section{Methodology}
\label{S2:5}

This section describes the suggested estimation methodology to study multifractionality of the CMB data that is based on theoretical results from Section~\ref{S2:4}. This and the next section also provide a detailed justification of this methodology and its assumptions and required modifications of the formulas for the spherical case and CMB data. 

For multifractional data, $H(t)$ changes from location to location and $H(t) \not\equiv$ const, where $t \in s_2(1)$. Several methods to estimate the local H\"older exponent are available in the literature. Different methods often give different results, see, for example, discussions in~\cite{Bianchi:2005, Struzik:2000} regarding inconsistent estimation results of the H\"older exponent. Inconsistent results by different techniques are due to their different assumptions, see~\cite{Bianchi:2005}. We propose an estimation method based on the generalized quadratic variations given by~\eqref{eq:2b} and~\eqref{eq:3b} and their asymptotic behaviour in~\eqref{eq:4b}. The results of this method are also compared with another conventional method that uses the rescale range (R/S) to estimate the H\"older exponent. This method is realized in the R package~{\sc{fractal}} \cite{fractal:2017}.

\begin{figure}[!htb]
\centering
\vspace{-0.3cm}
    \subfloat[HEALPix ring ordering]{\label{fig21a}
\includegraphics[trim={0cm 0cm 0cm 0cm},clip,width=0.35\textwidth, height=0.25\textheight]{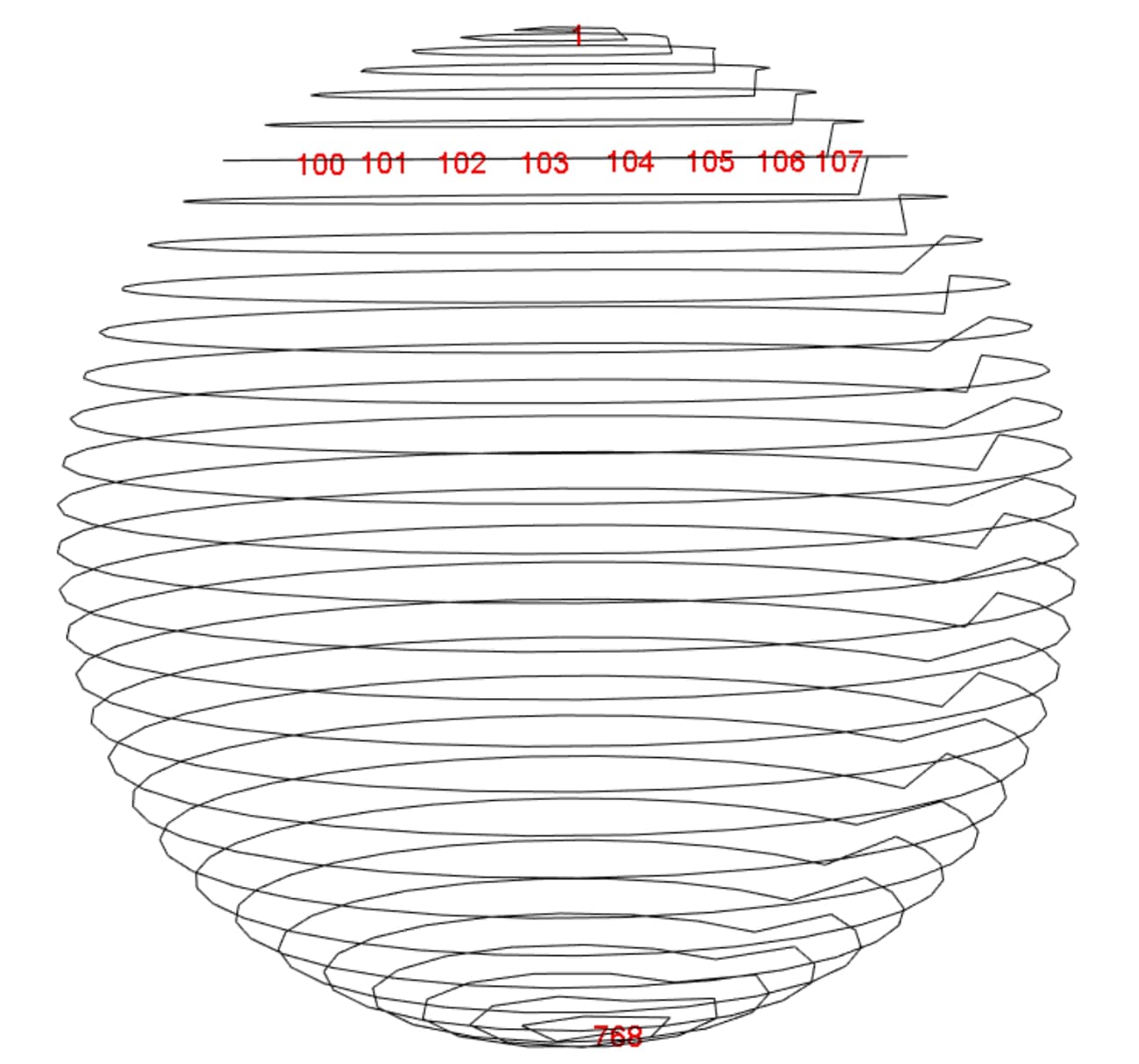}}
\centering
\vspace{-0.3cm}
    \subfloat[HEALPix nested ordering]{\label{fig21b}
\includegraphics[trim={0.1cm 1cm 0cm 0.1cm},clip,width=0.35\textwidth, height=0.25\textheight]{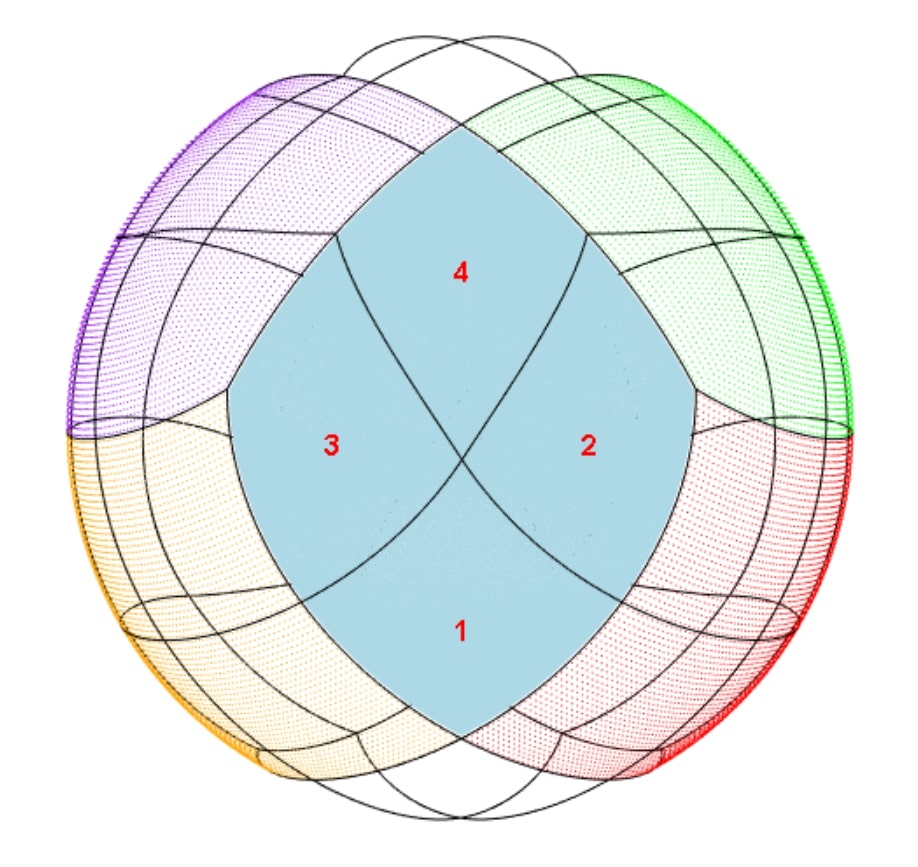}}
\vspace{0.4cm}
\caption{HEALPix ordering schemes}\label{fig21}
\end{figure}

Estimates of pointwise H\"older exponent values were computed using one- and two-dimensional regions of the the CMB data and the HEALPix ring and nested orderings~\cite{Gorski:2005}. These HEALPix ordering schemes are shown in Figure~\ref{fig21}. For the both cases, the highest available resolution, $N_{side}=2048$ was used.

The CMB data exhibit variations of the temperature intensities at very small scales ($\pm$ $1.8557 \times 10^{-3}$). To get reliable estimates of $H(t)$, a large amount of observations in neighbourhoods of each $t$ is required. Thus, in this publication, we do not discuss the preciseness of the local estimators of $H(t)$, but only pay attention to differences in the estimated values at different locations.

For computing purposes, the temperature intensities were scaled as \[\text{Scaled\ Intensity} (t) =\frac{\text{Intensity}(t)}{\max_{s\in s_2(1)}|\text{Intensity} (s)|}.\] It is clear from Definition~\ref{def2.4.1} that this scaling does not change the values of $\alpha_X(t)$. Also, by~\eqref{eq:2b} and~\eqref{eq:3b} the generalized quadratic variation of the scaled process $cX(t)$ is $c^2V_N^{(d)}(t)$, $d=1,2$. By~\eqref{eq:4b}, 
\begin{equation}
    \lim_{N \to \infty} \frac{\mathrm{log}\left(c^2V_{N}^{(d)}(t)\right)}{\mathrm{log}(N)} = \lim_{N \to \infty}\left(\frac{\mathrm{log}(c^2)}{\mathrm{log}(N)} + \frac{\mathrm{log}\Big(V_{N}^{(d)}(t)\Big)}{\mathrm{log}(N)}\right) = \lim_{N \to \infty} \frac{\mathrm{log}\Big(V_{N}^{(d)}(t)\Big)}{\mathrm{log}(N)},\label{eq:5b}
\end{equation}
which means that this scaling also does not affect $H(t)$.

As it was mentioned before, for small values of $\log(N)$ the estimates of $H(t)$ can be biased, which is now evident by the term $\frac{\mathrm{log} (c^2)}{\mathrm{log}(N)}$ in~\eqref{eq:5b}. However, this bias is due to the scaling effect only and is exactly the same for all values of $t$. Even if it might result in some errors in estimates $\hat{H}(t)$, it will not effect the analysis of differences in $H(t)$ values for different locations, which is the main aim of this analysis.

\section{Numerical studies}
\label{S2:6}

This section presents numerical studies and applications of the estimation methodology from Section~\ref{S2:5} to CMB data. The pointwise H\"older exponent estimates $\hat{H}(t)$ are computed and analysed for one- and two-dimensional regions of CMB data acquired from the NASA/IPAC Infrared Science Archive~{\cite{NASA:2019}}. The estimated H\"older exponents are used to quantify roughness of the CMB data. The developed methodology is also applied to detect possible anomalies in the CMB data.

\subsection{Estimates of H\"older exponent for one-dimensional CMB regions}
\label{S2:61}

For the one-dimensional case, the HEALPix ring ordered CMB temperature intensities were modelled by a stochastic process $X(t)$. Their H\"older exponents $H(t)$ were estimated by using the expression from the equation~\eqref{eq:4b} for the given large $N$ with $d=1$, where $V_{N}^{(1)}(t)$ was computed using the equation~\eqref{eq:2b}, which can be explicitly written as
\[V_N^{(1)}(t) = \sum_{p = 0}^{N-2}\left( X\left(\frac{p}{N}\right)- 2X\left(\frac{p+1}{N}\right)+ X\left(\frac{p+2}{N}\right)\right)^2.\]

As pixels on relatively small ring segments can be considered lying on approximately straight lines, the results from the case $d=1$ can be used. The parameter $N$ was chosen to give approximately the number of pixels within a half ring of the CMB sky sphere. The parameter $r$ is the distance from a HEALPix point $t$ that is the center of an interval in which we compute the total increment $V_{N}^{(1)}(t)$. By the expression of $v_N(t)$ in Definition~\ref{def2.4.2}, the parameter $\gamma$ was computed according to the formula, $\gamma={-\left(\log(r)/\log(N)\right)}$ for selected values of $N$ and $r$. Then, it was used in the equation~\eqref{eq:4b} to compute the estimated pointwise H\"older exponent values. 

According to the HEALPix structure of the CMB data with the resolution $N_{side}=2048$, there are $50331648$ pixels on the CMB sky sphere. The HEALPix ring ordering scheme results in ${4 \times N_{side}-1}$ rings, see~\cite{HEALPix:2020}. That is, for $N_{side} = 2048$, the CMB sky sphere consists of $8191$ rings. Based on the HEALPix geometry, the number of pixels in the upper part rings increase with the ring number, $\text{Ring}=1,...,2047$, as $(4 \times \text{Ring})$. The $(2{N_{side}}+1)=4097$ set of rings in the middle part of the CMB sky sphere have equal number of pixels which is $4 \times N_{side}$. The number of pixels in each of the final $(N_{side}-1)=2047$ rings in the lower part decreases according to the formula $(4 \times (8191-\text{Ring}+1))$.

For the one-dimensional case, the estimated pointwise H\"older exponent values $\hat{H}(t)$ were computed as follows. First, a random CMB pixel was selected and its ring was determined. Then pixels belonging to the half of that particular ring were selected. Then, for each CMB pixel in this rim segment, the quadratic variation was computed by $V_{N}^{(1)}(t)$ given in equation~\eqref{eq:2b}. When computing the generalized quadratic variation for a CMB pixel, the pixels within a distance $r=0.08$ from it were considered. For these pixels, the squared increments were computed and used to obtain the total of increments. Finally, the H\"older exponents were estimated by substituting the total of increments and the other parameters in the equation~\eqref{eq:4b}.

First, three CMB pixels {\lq552300\rq}, {\lq1533000\rq}, {\lq3253800\rq} located in the corresponding upper part rings 525, 875 and 1275 were chosen. Then for each CMB pixel in these half rings, their corresponding estimated H\"older exponents $\hat{H}(t)$ were computed. Next, another three pixels {\lq10047488\rq}, {\lq32575488\rq}, {\lq39948288\rq} were chosen in the middle part of the CMB sky sphere. Their ring numbers were 2250, 5000 and 5900~respectively. Finally, three CMB pixels {\lq47656664\rq}, {\lq48651704\rq}, {\lq49375304\rq} belonging to the lower part rings, 7035, 7275 and 7500 were selected and the pointwise H\"older exponents of pixels in their rim segments were estimated. 

\begin{figure}[!htb]
    \centering  
    \subfloat[Scaled intensities of ring~1275]{\label{fig22a}
    \includegraphics[trim={0cm 0cm 0cm 0cm},clip,width=0.35\textwidth, height=0.26\textheight]{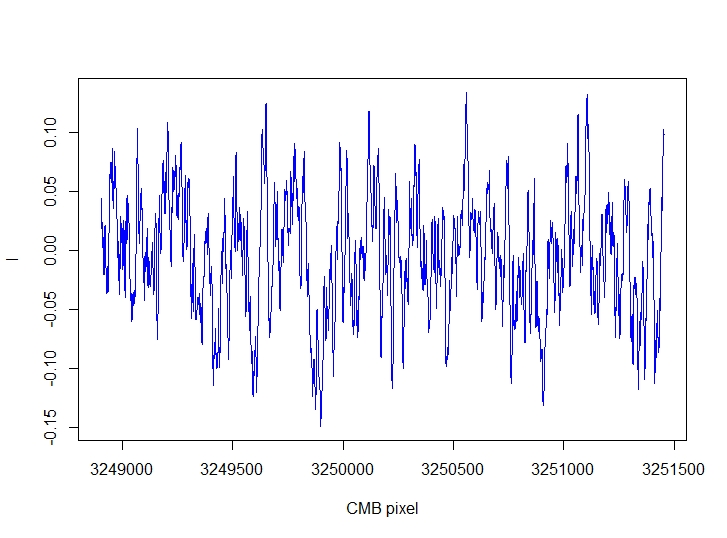}}            
    \centering
    \subfloat[Scaled intensities of ring~5900]{\label{fig22b}
    \includegraphics[trim={0cm 0cm 0cm 0cm},clip,width=0.35\textwidth, height=0.26\textheight]{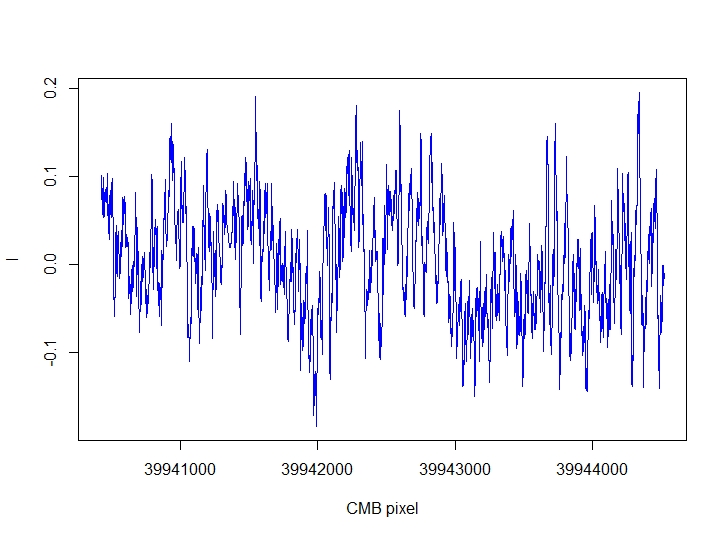}}\\
    \centering 
    \subfloat[$\hat{H}(t)$ values of ring~1275]{\label{fig22c}
    \includegraphics[trim={0cm 0cm 0cm 0cm},clip,width=0.35\textwidth, height=0.26\textheight]{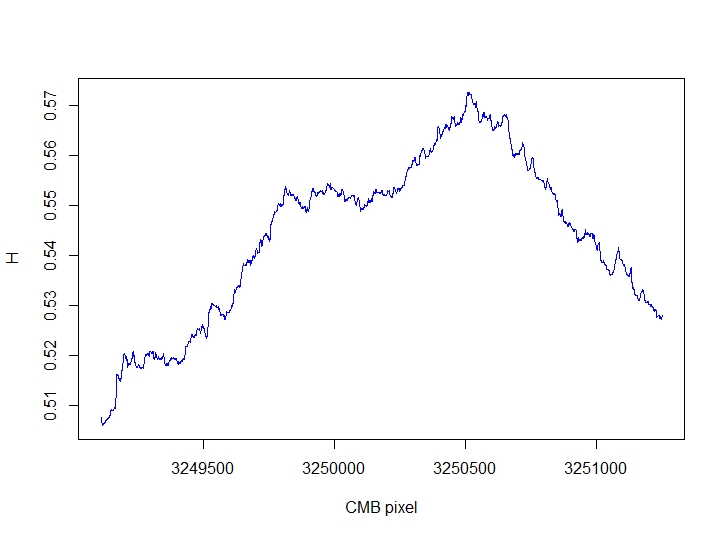}}
    \centering 
    \subfloat[$\hat{H}(t)$ values of ring~5900]{\label{fig22d}
    \includegraphics[trim={0cm 0cm 0cm 0cm},clip,width=0.35\textwidth, height=0.26\textheight]{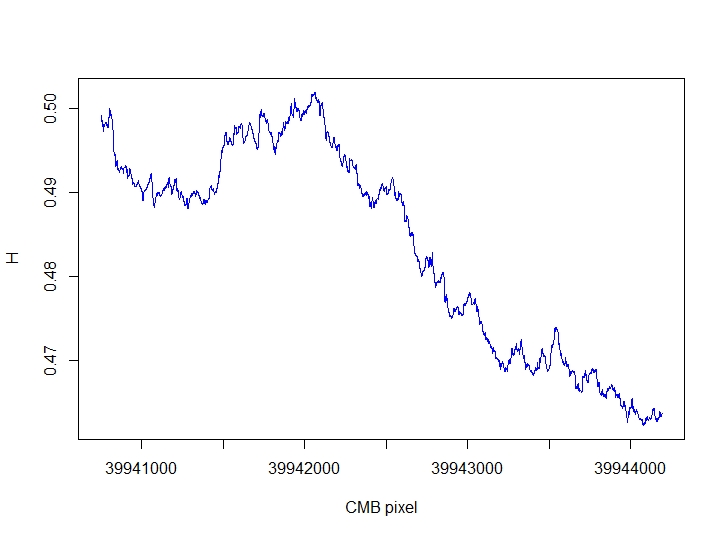}}
    \caption{Examples of scaled intensities and $\hat{H}(t)$ values for one-dimensional CMB regions}
    \label{fig22}
\end{figure}

For example, Figure~\ref{fig22} shows the plots of the scaled intensities and the estimated pointwise H\"older exponents of the rim segments of rings 1275 and 5900, which belong to the upper and middle parts of the CMB sky sphere respectively. It can be seen from Figures~\ref{fig22a} and~\ref{fig22b} that the majority of scaled intensities fall into the range $[-0.2, 0.2]$ and their fluctuations are random. Figures~\ref{fig22c} and~\ref{fig22d} exhibit that the $\hat{H}(t)$ values in both rim sections are changing and the dispersion range for ring~1275 is wider than that of ring~5900. Similar plots and results were also obtained for other rings.

\begin{table}[htb!]
\setlength{\tabcolsep}{4pt}
\centering
\small
\tabcolsep=0.11cm
\begin{tabular}{@{}*{7}{c}@{}}
\hline
\textbf{\thead{Part of CMB sky}} & \textbf{\thead{Case}} & \textbf{\thead{Ring number}} & \textbf{${\gamma}$} & \textbf{[${\hat{H}(t)}_{\min}$, ${\hat{H}(t)}_{\max}$]} & \textbf{${\hat{H}(t)}_{\max}$-${\hat{H}(t)}_{\min}$} & \textbf{\thead{Mean $\hat{H}(t)$}}\\
\hline
\multirow{3}{4em}{Upper part} & 1 & 525 & {0.3631} & {[0.5681, 0.6215]} & {0.0534} & {0.5960} \\ 
& 2 & 875 & {0.3382} & {[0.5443, 0.5782]} & {0.0339} & {0.5605} \\ 
& 3 & 1275 & {0.3220} & {[0.5059, 0.5727]} & {0.0668} & {0.5439} \\ 
\hline
\multirow{3}{4em}{Middle part} & 4 & 2250 & {0.3037} & {[0.4824, 0.5479]} & {0.0655} & {0.5137} \\ 
& 5 & 5000 & {0.3037} & {[0.4372, 0.4847]} & {0.0475} & {0.4626} \\ 
& 6 & 5900 & {0.3037} & {[0.4622, 0.5019]} & {0.0397} & {0.4835} \\ 
\hline
\multirow{3}{4em}{Lower part} & 7 & 7035 & {0.3260} & {[0.5067, 0.5384]} & {0.0317} & {0.5234} \\ 
& 8 & 7275 & {0.3361} & {[0.5256, 0.5553]} & {0.0297} & {0.5410} \\ 
& 9 & 7500 & {0.3492} & {[0.5548, 0.5896]} & {0.0348} & {0.5701} \\ 
\hline
\end{tabular}
\caption{Summary of $\hat{H}(t)$ values for pixels in different rings of the CMB sky sphere}\label{table21}
\end{table}

The summary of the estimated pointwise H\"older exponent values obtained by the discussed methodology is shown in Table~\ref{table21}. It is clear that the dispersion range of the $\hat{H}(t)$ values and the mean $\hat{H}(t)$ value change with ring numbers. These results suggest that the pointwise H\"older exponent values change from location to location.
The summary of the estimated pointwise H\"older exponent values obtained by the conventional (R/S) method using the command \enquote{RoverS} from the R package~{\sc{fractal}} is given in Table~\ref{table22}. It can be seen that the dispersion range and the mean $\hat{H}(t)$ value change with the spiraling ring number. Similar results were also obtained for other available estimators of the H\"older exponent. Although these numerical values are inconsistent between different methods, they all suggest that the pointwise H\"older exponent values change from location to location.

\begin{table}[!htb]
\setlength{\tabcolsep}{4pt}
\centering
\small
\tabcolsep=0.11cm
\begin{tabular}{@{}*{7}{c}@{}}
\hline
\textbf{\thead{Part of CMB sky}} & \textbf{\thead{Case}} & \textbf{\thead{Ring number}} &  \textbf{[${\hat{H}(t)}_{\min}$, ${\hat{H}(t)}_{\max}$]} & \textbf{${\hat{H}(t)}_{\max}$-${\hat{H}(t)}_{\min}$} & \textbf{\thead{Mean $\hat{H}(t)$}}\\
\hline
\multirow{3}{4em}{Upper part} & 1 & 525 & {[0.8106, 0.9035]} & {0.0929} & {0.8758} \\ 
& 2 & 875 & {[0.8527, 0.9146]} & {0.0619} & {0.8867} \\ 
& 3 & 1275 & {[0.8577, 0.9088]} & {0.0511} & {0.8898} \\ 
\hline
\multirow{3}{4em}{Middle part} & 4 & 2250 & {[0.8757, 0.9148]} & {0.0391} & {0.8975} \\ 
& 5 & 5000 & {[0.8656, 0.9079]} & {0.0423} & {0.8883} \\ 
& 6 & 5900 & {[0.8702, 0.9072]} & {0.0370} & {0.8926} \\ 
\hline
\multirow{3}{4em}{Lower part} & 7 & 7035 & {[0.8617, 0.9081]} & {0.0464} & {0.8889} \\ 
& 8 & 7275 & {[0.8599, 0.9098]} & {0.0499} & {0.8897} \\ 
& 9 & 7500 & {[0.8348, 0.9004]} & {0.0656} & {0.8714} \\ 
\hline
\end{tabular}
\caption{Summary of $\hat{H}(t)$ values for pixels in different rings of the CMB sky sphere using the R/S method}\label{table22}
\end{table}

It is expected that temperature intensities are positively dependent/correlated in close regions, see the covariance analysis in~\cite{Broadbridge:2019}. Therefore, running standard equality of means tests under independence assumptions will provide even more significant results if the hypothesis of equal means is rejected.

To prove that distributions of $\hat{H}(t)$ are statistically different between different sky regions, we carried out several equality of means tests. Before that, the Shapiro test was used to ensure that the $\hat{H}(t)$ values satisfy the normality assumption. For all the considered cases in Table~\ref{table21}, their $\hat{H}(t)$ values failed the normality assumption. Since the CMB pixels close to each other can be dependent, to get more reliable results we chose CMB pixels at distance~50 apart on a ring. The Shapiro test confirmed that in all the considered upper and lower part cases in Table~\ref{table21}, $\hat{H}(t)$ values at step~50 satisfied the normality assumption, whereas the $\hat{H}(t)$ values in the middle part failed the normality assumption.   

\begin{figure}[!htb]
\centering
\RawFloats
  \begin{minipage}[b]{0.5\linewidth}
    \centering
    \includegraphics[trim={0cm 0cm 0.1cm 0cm},clip,width=0.75\linewidth, height=0.5\linewidth]{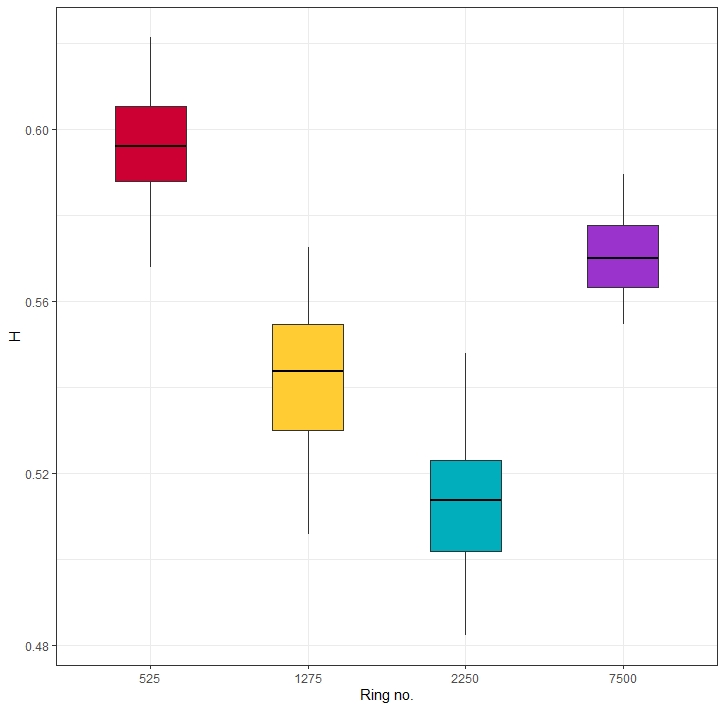}
    \captionof{figure}{The distribution of $\hat{H}(t)$ values of four rim segments}\label{fig23}
  \end{minipage}%
  \begin{minipage}[b]{0.5\linewidth}
    \small
    \centering
\begin{tabular}[b]{|*{4}{c|}}
                                                                     \cline{2-2}
  \multicolumn{1}{l|}{1275} & {$3.048 \times 10^{-15}$}                 \\ \cline{2-3}
  \multicolumn{1}{l|}{2250} & {$7.939 \times 10^{-11}$} & {$3.606 \times 10^{-12}$}             \\ \cline{2-4}
  \multicolumn{1}{l|}{7500} & {$1.533 \times 10^{-8}$} & {$4.605 \times 10^{-10}$} & {$3.717 \times 10^{-13}$}        \\ \cline{2-4}
  \multicolumn{1}{c}{} & \multicolumn{1}{c}{525} & \multicolumn{1}{c}{1275} & \multicolumn{1}{c}{2250} 
\end{tabular}
\captionof{table}{p-values for Wilcoxon tests between different rings}\label{table23}
\end{minipage}
\end{figure}

Let $\mu_1$ and $\mu_2$ be the $\text{mean}{({\hat{H}(t)})}$ values of the rim segments of rings 525 and 1275 respectively. To test the hypothesis $H_0: \mu_1 = \mu_2$ vs. $H_1: \mu_1 \neq \mu_2$ we carried out the Wilcoxon test. The obtained p-value ($3.048 \times 10^{-15}$) is significantly less than 0.05 and suggests that the means are different at {5\%} level of significance. Similar results were obtained for the Wilcoxon tests between all pairs of the cases in Table~\ref{table21}. For example, Table~\ref{table23} shows Wilcoxon test results for selected four rings, two in the upper part, and the other two correspondingly in the middle and lower parts of the CMB sky sphere. Figure~\ref{fig23} shows the distribution box plots of the $\hat{H}(t)$ values in the rim segments of rings 525, 1275, 2250 and 7500. It is clear from Figure~\ref{fig23} that the $\text{mean}{({\hat{H}(t)})}$ values are different from each other in these cases.

Analogously to Table~\ref{table23}, for all carried out Wilcoxon tests between the rim sectors in the upper, middle and lower parts, their p-values < 0.05. Therefore, there is enough statistical evidence to suggest that the pointwise H\"older exponents change from location to location. While we compared H\"older exponents for different rings, from Figure~\ref{fig22} it is clear that $\hat{H}(t)$ is also changing for pixels within same rings.

\subsection{Estimates of H\"older exponent for two-dimensional CMB regions}
\label{S2:62}

For two-dimensional sky regions, pointwise H\"older exponent values $H(t)$ were estimated according to the equation~\eqref{eq:4b} with $d=2$, where $V_{N}^{(2)}(t)$ was computed using the equation \eqref{eq:3b}. The equation \eqref{eq:3b} in Definition~\ref{def2.4.3} can be written in the following explicit form
\begin{multline*}
V_{N}^{(2)}(t) = \sum_{p \in v_{N}(t)}\left(\sum_{k_1 \in \{0,1,2\}} \sum_{k_2 \in \{0,1,2\}} e_{k_1}e_{k_2} X\left(\frac{p_1+k_1}{N},\frac{p_2+k_2}{N}\right)\right)^2
= \sum_{p \in v_{N}(t)} \bigg(X\left(\frac{p_1}{N},\frac{p_2}{N}\right)\\ - 2X\left(\frac{p_1}{N},\frac{p_2+1}{N}\right) -2 X\left(\frac{p_1+1}{N},\frac{p_2}{N}\right) +  X\left(\frac{p_1}{N},\frac{p_2+2}{N}\right) +  X\left(\frac{p_1+2}{N},\frac{p_2}{N}\right) + 4 X\left(\frac{p_1+1}{N},\frac{p_2+1}{N}\right) \\ -2 X\left(\frac{p_1+1}{N},\frac{p_2+2}{N}\right) -2 X\left(\frac{p_1+2}{N},\frac{p_2+1}{N}\right) +  X\left(\frac{p_1+2}{N},\frac{p_2+2}{N}\right)\bigg)^2.
\end{multline*}

\begin{figure}[!htb]
\centering
\vspace{-0.3cm}
\includegraphics[trim={0.7cm 2.2cm 0.7cm 2.6cm},clip,width=0.35\textwidth, height=0.22\textheight]{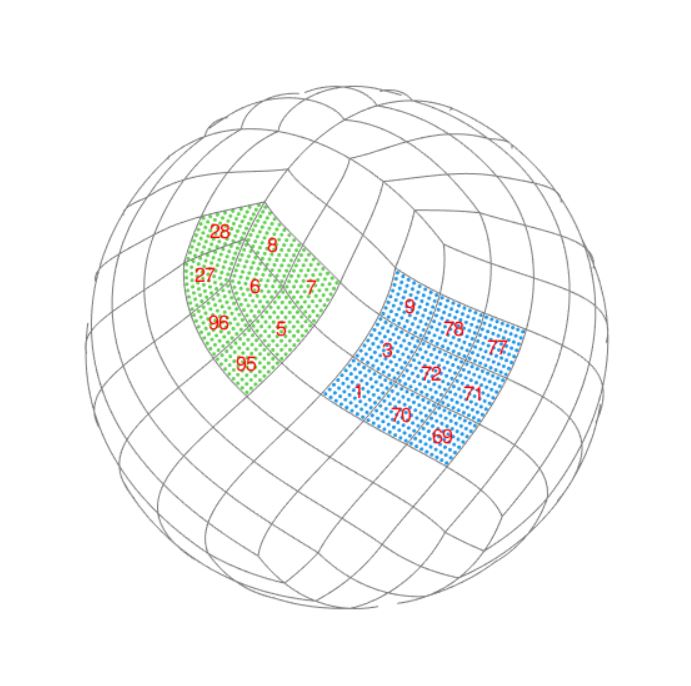}
\vspace{0cm}
\caption{Examples of pixels with 7 and 8 neighbours for $N_{side}=4$}\label{fig24}
\end{figure}

To compute quadratic increments of spherical random fields, relatively small parts of the sphere can be approximately considered as regions of the plane and the above formula can be applied. Note that the internal summation set $\left\{ \left( {\frac{p_1 + k_1}{N}}, {\frac{p_2 + k_2}{N}}\right): k_1, k_2 \in \{0, 1, 2\} \right\}$ can be very efficiently represented by the HEALPix nested structure. Indeed, all pixels have either 7 or 8 neighbours, see~Figure~\ref{fig24}. The $3 \times 3$ configuration with 8 neighbours perfectly matches the internal summation set and can be directly used in computations of $V_N^2(t)$. For the case of 7 neighbours, an additional $8^{th}$ neighbour which intensity equals to the one of its adjusted pixel was imputed. For the resolution $N_{side} = 2048$ only 24 out of 50331648 pixels have 7 neighbours. For such small number of pixels the imputation  has a negligible impact on the results. 

Circular regions with radius $R=0.23$ were used in this section computations. Let $N$ denote the number of pixels within such circular regions. Then, $N \approx$ 662700 pixels. To reduce the computation time, we chose a grid of~1000 CMB pixels with the step 662 = [662700/1000], where $[\cdot]$ denotes the integer part, over the total number of pixels. To compute local estimators $\hat{H}(t)$, for each chosen CMB pixel, a circular window with radius $r=0.01$ was selected. The value of $\gamma$ was computed as $\gamma={-\left(\log(\sqrt{\pi}r/2)/\log(\sqrt{N})\right)}$ for given values of $N$ and $r$. The factor $\sqrt{\pi}/2$ appeared as the number of pixels is proportional to a window area. To match the number of pixels in circular window regions that were used in computations and square regions used for summation in $V_{N}^{(2)}(t)$, the length $2{d_0}$ of the side of squares should satisfy the equation ${(2{d_0})^2} = {\pi r^2}$. The obtained $\gamma$ was substituted in the equation \eqref{eq:4b} to compute the estimated pointwise H\"older exponent values. For $r=0.01$, there are approximately~2836 pixels in each specified window. For each of these pixels, the squared increment was computed and the total of increments was obtained by the expression for~$V_{N}^{(2)}(t)$.\vspace{-5mm}
\begin{figure}[!htb]
    \centering  
    \subfloat[A sky window from the warm region]{\label{fig25a}
    \includegraphics[trim={0cm 0.5cm 0cm 0cm},clip,width=0.35\textwidth, height=0.26\textheight]{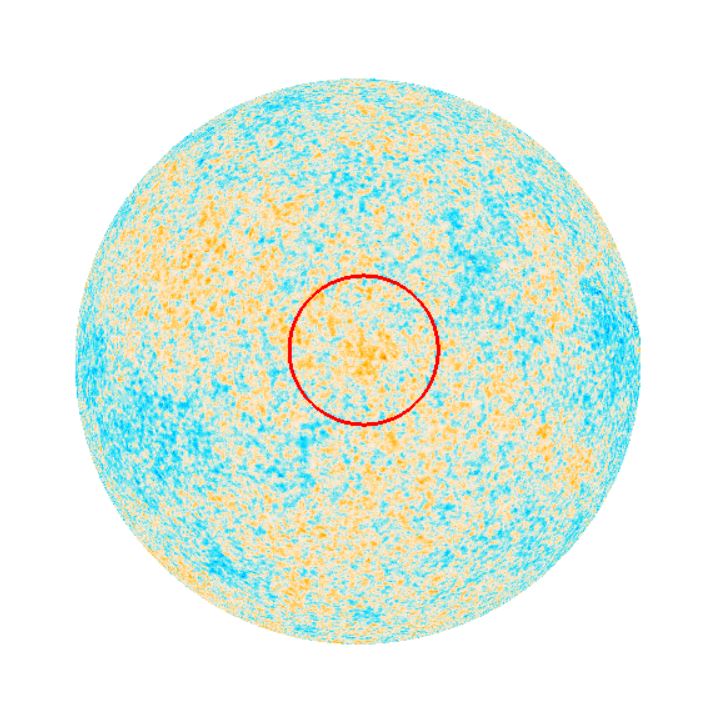}}            
    \centering
    \subfloat[A sky window from the cold region]{\label{fig25b}
    \includegraphics[trim={0cm 0.5cm 0cm 0cm},clip,width=0.35\textwidth, height=0.26\textheight]{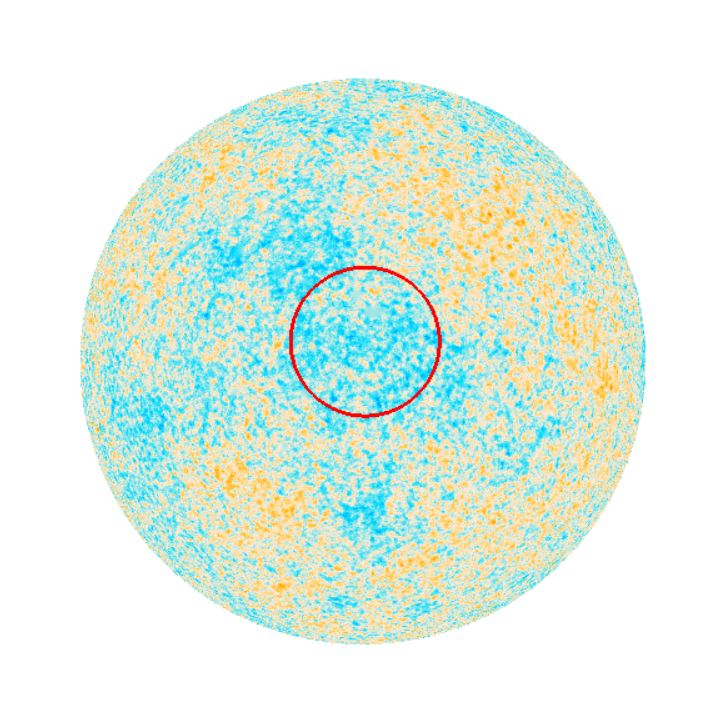}}\\
    \centering 
    \subfloat[A sky window with a mixture of temperatures]{\label{fig25c}
    \includegraphics[trim={0cm 0.5cm 0cm 0cm},clip,width=0.35\textwidth, height=0.26\textheight]{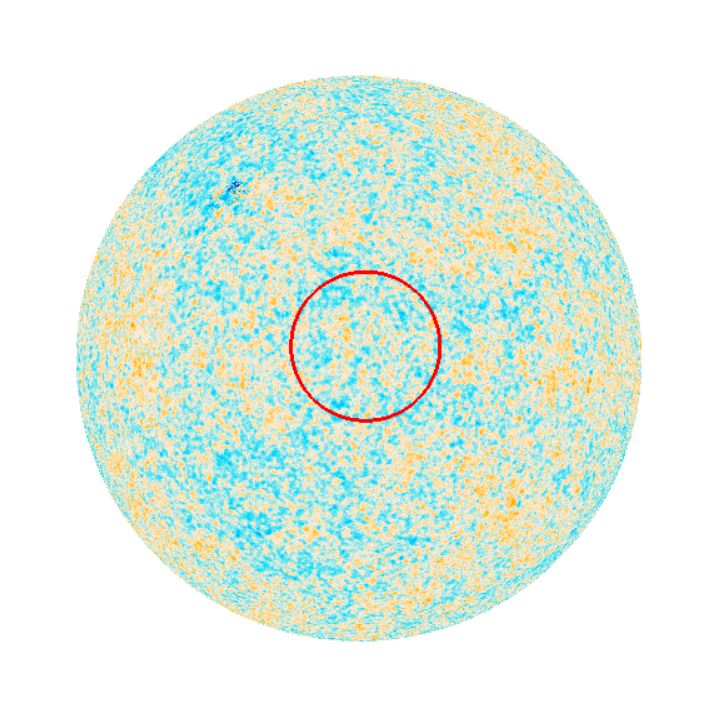}}
    \centering 
    \subfloat[A sky window with the borderline region]{\label{fig25d}
    \includegraphics[trim={0cm 0.5cm 0cm 0cm},clip,width=0.35\textwidth, height=0.26\textheight]{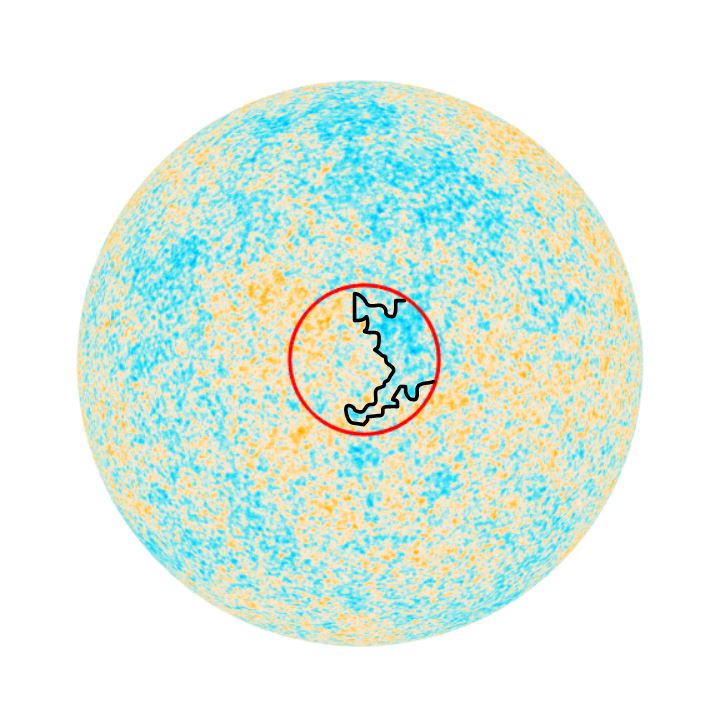}}
    \caption{Sky windows used for computations}
    \label{fig25}
\end{figure}

First, a circular CMB sky window of radius $R=0.23$ from a warm area with the majority of high temperature intensities was selected. The mean temperature intensity in the selected CMB sky region covering the warm area was $5.97861 \times 10^{-5}$. The window is shown in Figure~\ref{fig25a}. The number of pixels in that specific window was~662685. Then, different circular CMB sky windows having a radius of $R=0.23$ covering cold, mixture of warm and cold regions and having a borderline of warm and cold regions shown in Figures~\ref{fig25b}, \ref{fig25c}, and~\ref{fig25d} were chosen. In each of the cold, mixture of warm and cold and a borderline having warm and cold regions, the number of pixels were~662697, 662706 and 662725 respectively. The value of $\gamma$ was computed as $\gamma=0.705$ for each CMB sky region. The corresponding mean temperature intensities were $-8.34055 \times 10^{-5}$, $-1.74035 \times 10^{-5}$ and $7.59851 \times 10^{-6}$.

\begin{figure}[!htb]
    \centering  
    \subfloat[$\hat{H}(t)$ values from the warm region]{\label{fig26a}
    \includegraphics[trim={0cm 0cm 0cm 0cm},clip,width=0.35\textwidth, height=0.26\textheight]{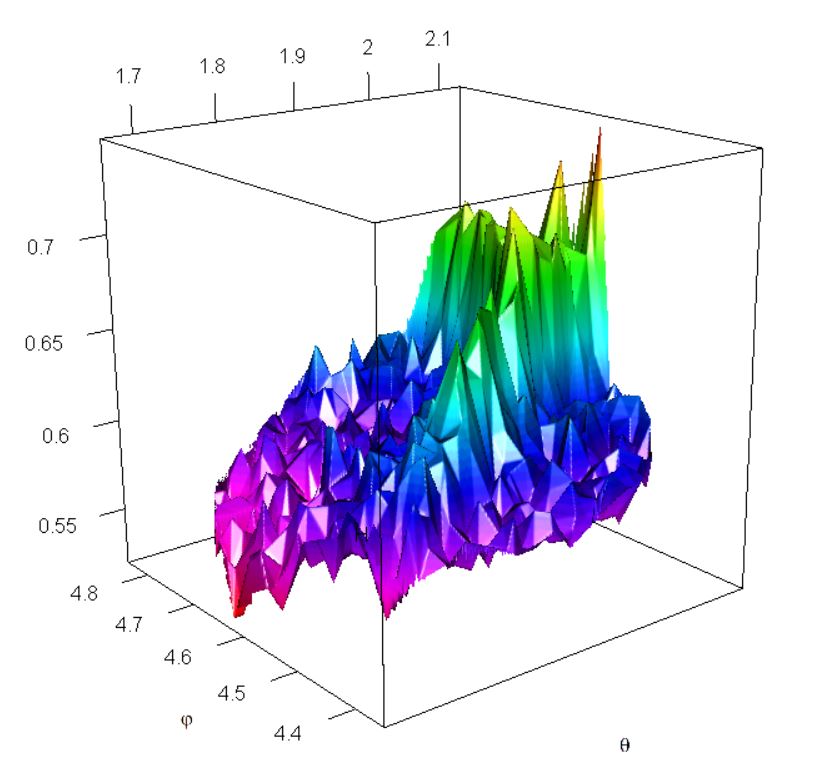}}            
    \centering
    \subfloat[$\hat{H}(t)$ values from the cold region]{\label{fig26b}
    \includegraphics[trim={0cm 0cm 0cm 0cm},clip,width=0.35\textwidth, height=0.26\textheight]{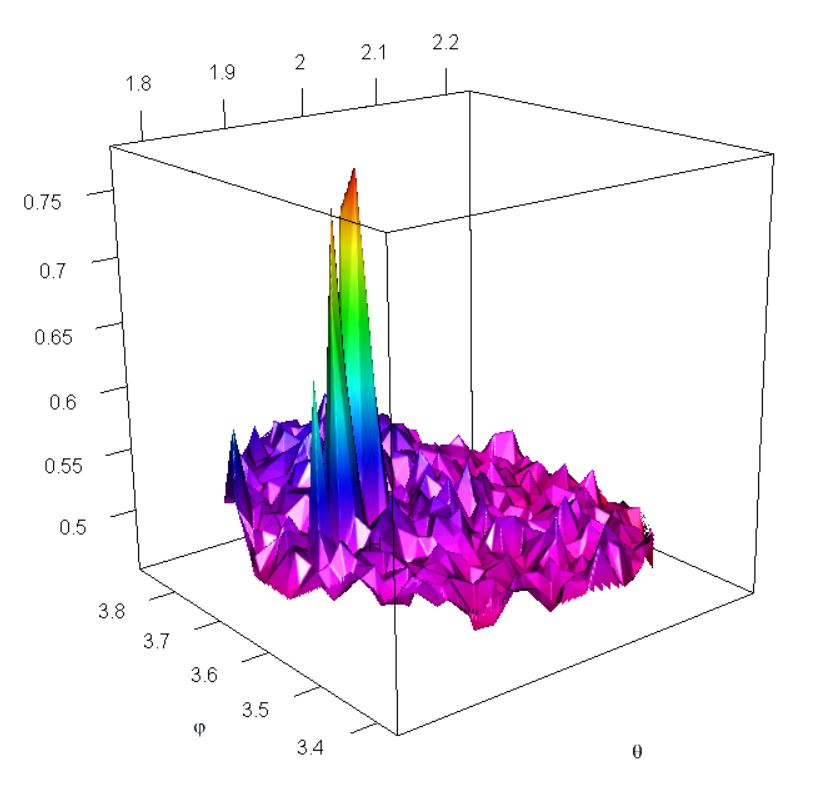}}\\
    \centering 
    \subfloat[$\hat{H}(t)$ values from the region with mixture of temperatures]{\label{fig26c}
    \includegraphics[trim={0cm 0cm 0cm 0cm},clip,width=0.35\textwidth, height=0.26\textheight]{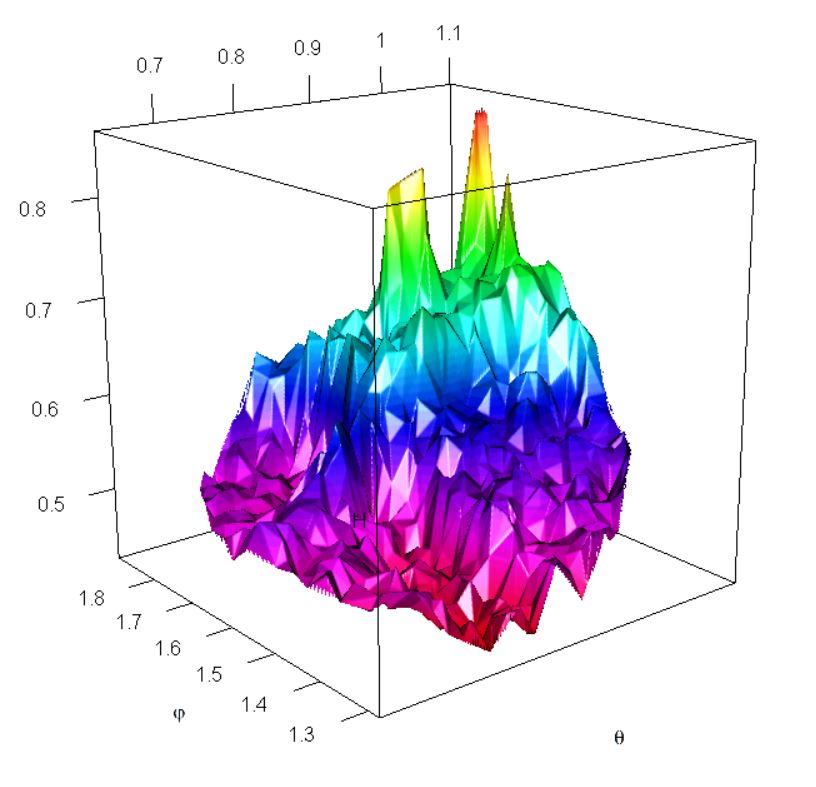}}
    \centering 
    \subfloat[$\hat{H}(t)$ values from the borderline region]{\label{fig26d}
    \includegraphics[trim={0cm 0cm 0cm 0cm},clip,width=0.35\textwidth, height=0.26\textheight]{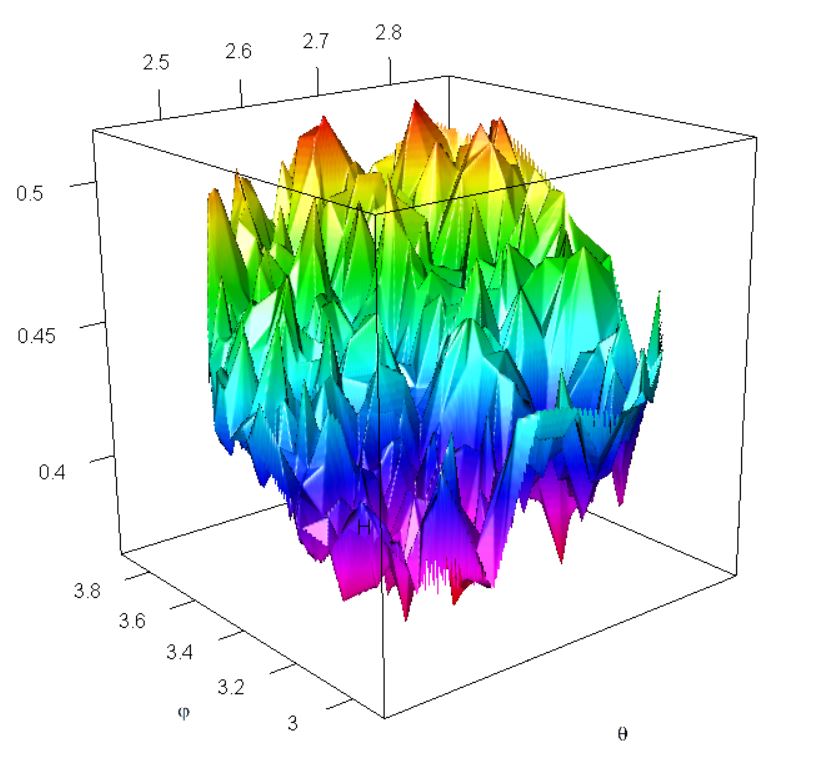}}
    \caption{Local estimates $\hat{H}(t)$ for two-dimensional regions}
    \label{fig26}
\end{figure}

The plots of the estimated pointwise H\"older exponent values for each case are displayed in Figures~\ref{fig26a}, \ref{fig26b}, \ref{fig26c} and \ref{fig26d}. These $\hat{H}(t)$ values are mostly dispersed in the interval $[0.36, 0.86]$. Figures~\ref{fig26a}, \ref{fig26b}, \ref{fig26c} and \ref{fig26d} show an erratic and an irregular behaviour in the distribution of $\hat{H}(t)$ values. It can be noticed that the estimates in Figures~\ref{fig26a} and \ref{fig26d} with substantial warm temperatures have larger $\hat{H}(t)$ fluctuations than the $\hat{H}(t)$ values for cold regions. 

The summary of the estimated pointwise H\"older exponents for each selected region is given in Table~\ref{table24}. It shows the mean CMB temperature intensities of each circular window. Table~\ref{table24} also presents the estimated minimum, maximum and mean $\hat{H}(t)$ values computed by using the selected 1000 CMB pixels. It is clear from Table~\ref{table24} that the mean $\hat{H}(t)$ value from the warm region is the highest and it is the lowest for the borderline region. The mean $\hat{H}(t)$ values of the cold region and mixture case lie in between them. It is apparent from Table~\ref{table24} that the range of the estimated pointwise H\"older exponent values change with respect to the temperature of the chosen regions of the CMB sky sphere. 

To further investigate the estimated pointwise H\"older exponents, they were computed for 100 random CMB pixels in each of the considered regions. It was apparent that even if one accounts for variation by considering these 100 CMB pixels, the $\hat{H}(t)$ values between different regions are different. The analyses suggested that all $\hat{H}(t)$ values for 100 and 1000 CMB pixels are consistent. Therefore, the results suggest that the estimated pointwise H\"older exponent values change from place to~place.

\begin{table}[h!]
\setlength{\tabcolsep}{4pt}
\centering
\footnotesize
\tabcolsep=0.11cm
\begin{tabular}{@{}*{7}{c}@{}}
\hline
\textbf{\thead{Inspection\\ Window}} & \textbf{\thead{Mean Intensity}}  & \textbf{[${\hat{H}(t)}_{\min}$, ${\hat{H}(t)}_{\max}$]} & \textbf{${\hat{H}(t)}_{\min}$-${\hat{H}(t)}_{\max}$} & \textbf{\thead{Mean $\hat{H}(t)$}}\\
\hline
\textbf{Warm region} & {\SI{5.97861e-05}{}} & {[0.5217, 0.7484]} & {0.2267} & {0.5994} \\
\textbf{Cold region} & {\SI{-8.34055e-05}{}} & {[0.4534, 0.7806]} & {0.3272} & {0.5151} \\
\textbf{Mixture case} & {\SI{-1.74035e-05}{}} & {[0.4302, 0.8592]} & {0.4290} & {0.5563} \\
\textbf{Borderline case} & {\SI{7.59851e-06}{}} & {[0.3629, 0.5158]} & {0.1529} & {0.4407} \\
\hline
\end{tabular}
\caption{Analysis of CMB sky windows with different temperatures}\label{table24}
\end{table}

\begin{figure}[!htb]
\centering
\RawFloats
  \begin{minipage}[b]{0.45\linewidth}
    \centering
    \includegraphics[trim={0cm 0cm 0.1cm 0cm},clip,width=0.8\linewidth, height=0.65\linewidth]{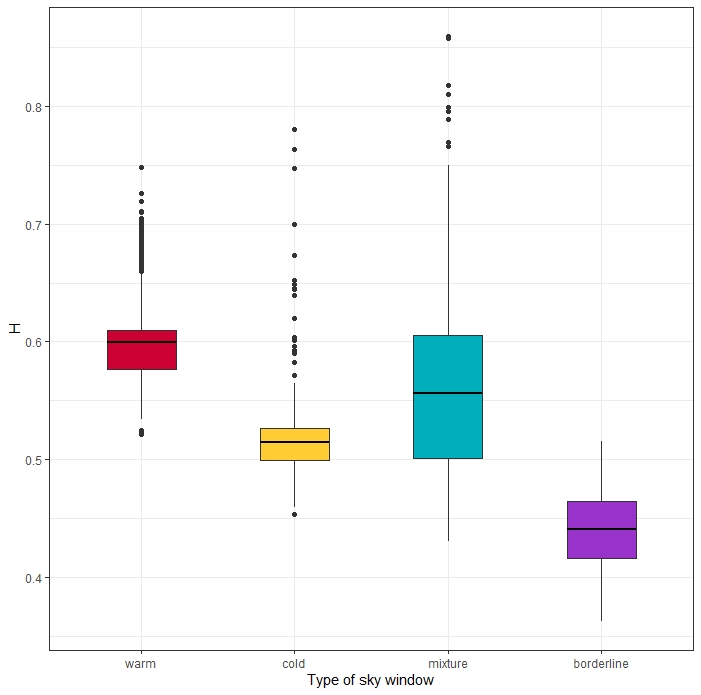}
    \captionof{figure}{The distribution of $\hat{H}(t)$ values for chosen\\ sky windows}\label{fig27}
  \end{minipage}%
  \begin{minipage}[b]{0.5\linewidth}
    \small
    \centering
\begin{tabular}[b]{|*{4}{c|}}
                                                                     \cline{2-2}
  \multicolumn{1}{l|}{cold} & {$< 2.2 \times 10^{-16}$}                 \\ \cline{2-3}
  \multicolumn{1}{l|}{mixture} & {$< 2.2 \times 10^{-16}$} & {$< 2.2 \times 10^{-16}$}             \\ \cline{2-4}
  \multicolumn{1}{l|}{borderline} & {$< 2.2 \times 10^{-16}$} & {$< 2.2 \times 10^{-16}$} & {$< 2.2 \times 10^{-16}$}        \\ \cline{2-4}
  \multicolumn{1}{c}{} & \multicolumn{1}{c}{warm} & \multicolumn{1}{c}{cold} & \multicolumn{1}{c}{mixture} 
\end{tabular}
\centering
\captionof{table}{p-values for Wilcoxon tests between chosen\\ sky windows}\label{table25}
\end{minipage}
\end{figure}

To prove that $\hat{H}(t)$ is statistically different between different sky windows, we carried out several equality of means tests. Initially, we carried out the Shapiro test to ensure that the $\hat{H}(t)$ values satisfy the normality assumption. However, for all the considered cases in Table~\ref{table24}, their $\hat{H}(t)$ values failed the normality assumption. Figure~\ref{fig27} displays the distribution box plots of the $\hat{H}(t)$ values in the CMB sky windows with warm, cold, mixture of temperatures and having a borderline region. It can be noticed from Figure~\ref{fig27} that the $\hat{H}(t)$ distributions have extreme values in all the four cases. Thus, we present only results from the Wilcoxon test as it is reliable amidst the non-normality of data and in the presence of outliers.

Let $\mu_1$ and $\mu_2$ be the $\text{mean}{({\hat{H}(t)})}$ values in the sky windows with warm and cold regions respectively. Testing the hypothesis $H_0: \mu_1 = \mu_2$ vs. $H_1: \mu_1 \neq \mu_2$ by carrying out the Wilcoxon test, we obtained a p-value ($< 2.2 \times 10^{-16}$) that is significantly less than 0.05. It suggests that the means are different at {5\%} level of significance. Similar results were obtained for the Wilcoxon tests between all pairs of the cases and the corresponding p-values are shown in Table~\ref{table25}. It suggests that the $\text{mean}{({\hat{H}(t)})}$ values are different from each other in all the cases. Apart from variations between cases, it can be observed from Figure~\ref{fig26} and Table~\ref{table24} that the estimated H\"older exponents do change within individual sky windows as well. 

Therefore, there is enough statistical evidence to suggest that the pointwise H\"older exponents change from location to location of the CMB sky sphere.

\subsection{Analysis of CMB temperature anomalies in the equatorial region of CMB sky sphere}
\label{S2:63}

As previously discussed in Section~\ref{S2:1}, several missions have measured the CMB temperature anisotropies gradually increasing their precision by using advanced radio telescopes. This section discusses applications of the multifractional methodology to detect regions of CMB maps with \enquote{anomalies}. In particular, it can help in evaluating various reconstruction methods for blocked regions with unavailable or too noisy~data.

It is well known that the CMB maps are affected by the interference coming from the Milky Way and radio signals emitting from our galaxy are much noisy than the CMB. Thus, the Milky Way blocks the CMB near the galactic plane. However, the smooth and predictable nature of Milky Way's radiation spectrum has enabled to disclose the cosmological attributes by subtracting the spectrum from the initially observed intensities~\cite{Castelvecchi:2019}. From Planck~2015 results, the CMB maps have been cleaned and reconstructed using different techniques namely, COMMANDER, NILC, SEVEM, SMICA  see~\cite{Adam:2016I, Ade:2016} for more information. We are using the CMB map produced from the SMICA method~\cite{NASA:2019} with $N_{side}=2048.$

To examine the random behaviour of isotropic Gaussian fields on the sphere, a direction dependent novel mathematical tool has been proposed in~\cite{Hamann:2021}. They have applied their probe to investigate the CMB maps from Planck PR2~2015 and PR3~2018 with specific consideration to cosmological data from the inpainted maps. To detect departures from the traditional statistical model of the CMB data, they have utilized the auto-correlation of the sequence of full-sky Fourier coefficients and have proposed an \enquote{AC discrepancy} function on the sphere. For the inpainted Planck~2015 COMMANDER map, \cite{Hamann:2021} shows the maximum \enquote{AC discrepancy} for the galactic coordinates\footnote{The galactic coordinate system with Sun as the center is used in astronomy to locate the relative positions of objects and motions within the Milky Way Galaxy. It consists of galactic longitude $l, 0 < l < 2\pi$ and galactic latitude $b, -\pi/2 < b < \pi/2$. They are related to the spherical coordinates by $l=\phi$ and $b=(\pi/2- \theta)$.}, $(l,b) = (353.54, 1.79)$. Similarly, for the inpainted Planck COMMANDER~2018, NILC~2018, SEVEM~2018 and SMICA~2018 with $N_{side}=1024$, there are significant departures at the galactic coordinates (12.57, 0.11), (61.17,-30.73), (261.25,-2.99) and (261.34,-2.99) respectively. A majority of these locations are the masked regions of the galactic plane. The galactic coordinates corresponding to the largest deviations are different for each map depicting the discrepancies in the underlying inpainting techniques.

The approach in~\cite{Hamann:2021} used directional dependencies in CMB data on the unit sphere. The results below are based on a different approach that uses the local roughness properties of these data. Therefore, the detected regions of high anomalies can be different for these two methods as they reflect different physical anisotropic properties of CMB, see, for example, Figure~\ref{fig210}. The estimated local H\"older exponents on one-dimensional rings can be considered as directional local probes of CMB anisotropy. However, the estimates for two-di\-mensional  regions are more complex and aggregate local information about roughness in different~directions.

\begin{figure}[!htb]
    \centering  
    \subfloat[Scaled intensities of great circle/ring 4096]{\label{fig28a}
    \includegraphics[trim={0cm 0cm 0cm 1.5cm},clip,width=0.35\textwidth, height=0.25\textheight]{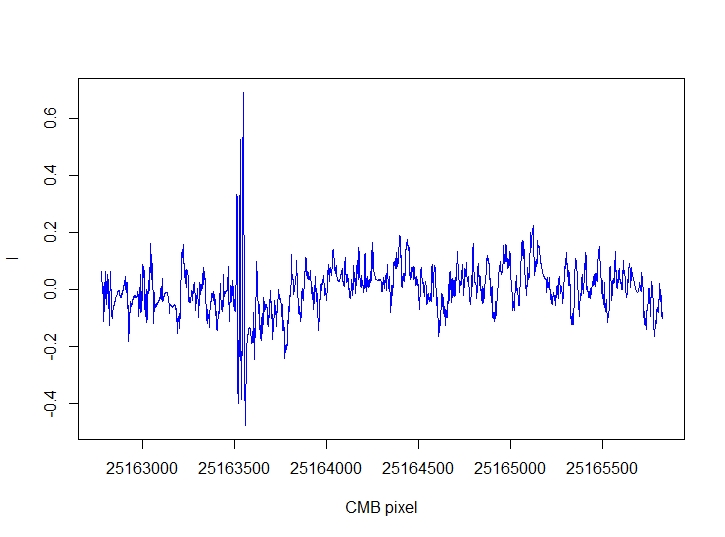}}            
    \centering
    \subfloat[$\hat{H}(t)$ values of great circle/ring 4096]{\label{fig28b}
    \includegraphics[trim={0cm 0cm 0cm 1.5cm},clip,width=0.35\textwidth, height=0.25\textheight]{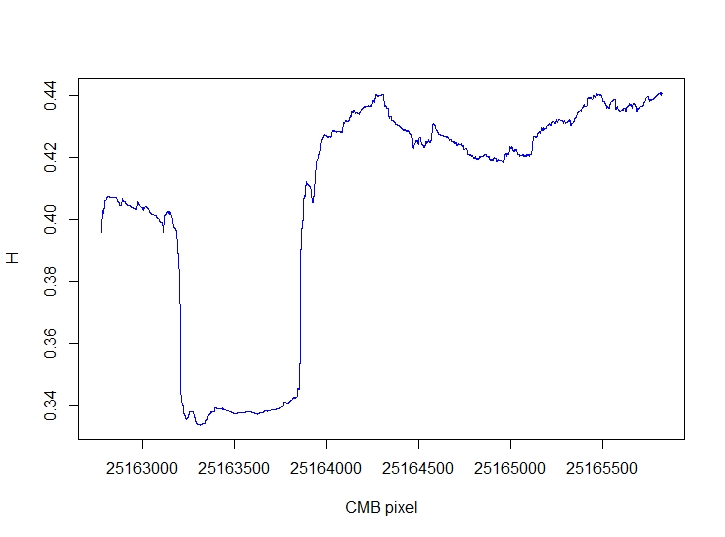}}\\
    \centering 
    \subfloat[Scaled intensities of equator region]{\label{fig28c}
    \includegraphics[trim={0cm 0cm 0cm 0cm},clip,width=0.35\textwidth, height=0.26\textheight]{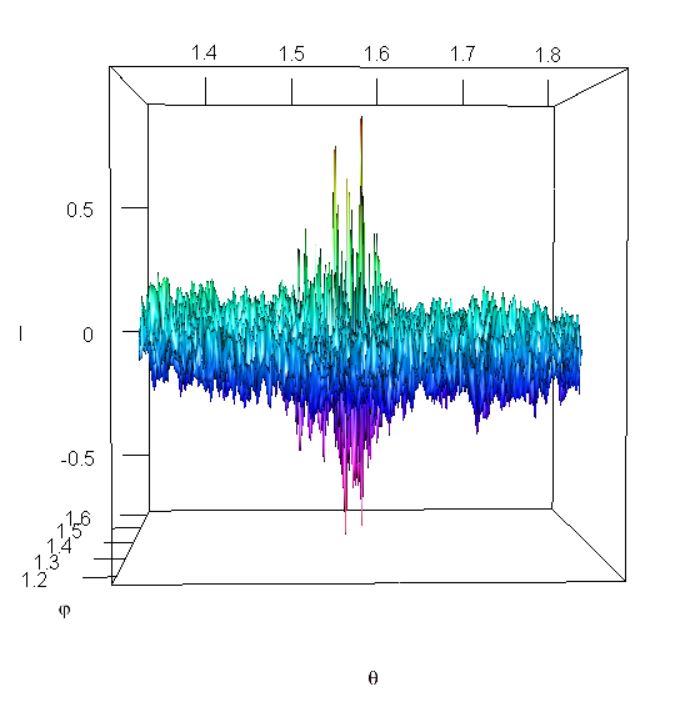}}
    \centering 
    \subfloat[$\hat{H}(t)$ values of equator region]{\label{fig28d}
    \includegraphics[trim={0cm 0cm 0cm 0cm},clip,width=0.35\textwidth, height=0.26\textheight]{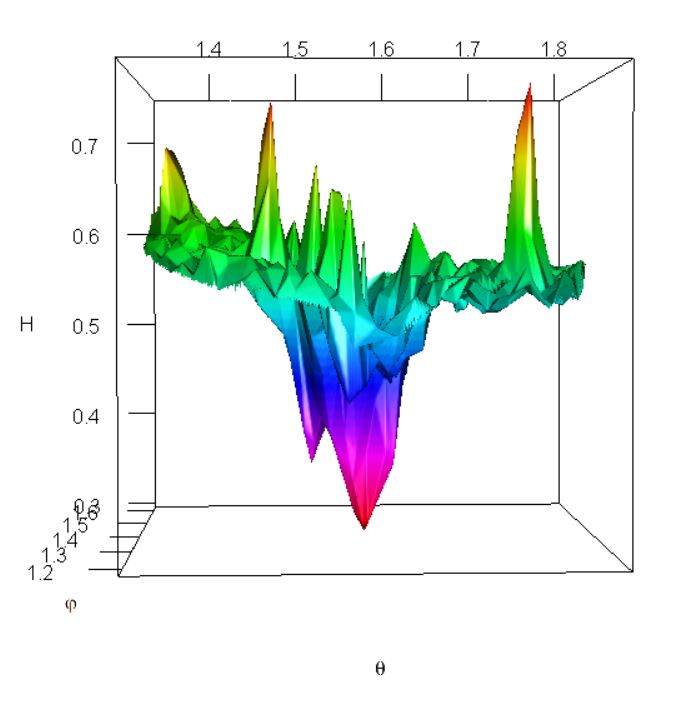}}
    \caption{Scaled intensities and estimated $\hat{H}(t)$ values in one- and two-dimensional regions of the great circle}
    \label{fig28}
\end{figure}

In the following analysis, we use estimated values of the H\"older exponent to detect regions of possible anomalies in CMB maps. Figure~\ref{fig28} shows the plots of scaled intensities and estimated H\"older exponent values $\hat{H}(t)$ in one- and two-dimensional CMB regions of the great circle. It can be noticed from Figure~\ref{fig28a} that there is an increase in the fluctuations of the scaled intensity values between the HEALPix range $[25163000, 25164000]$ of the great circle ring. A low plateau of estimated $\hat{H}(t)$ values in Figure~\ref{fig28b} corresponds to that range of HEALPix values. The equator rim segment with the unusual plateau of $\hat{H}(t)$ values has CMB pixel numbers ranging from 25163208 to 25163852. Their corresponding galactic coordinates were found to be between, $(65.02, 0.01)$ and $(93.32, 0.01)$.

Similarly, this unusual behaviour of $\hat{H}(t)$ values was observed in the two-dimensional CMB regions near the galactic plane/equator. Figure~\ref{fig28c} shows the plot of scaled intensities in the two-dimensional space and a spike in intensities can be observed near the specified range of HEALPix values. The corresponding lower valley of $\hat{H}(t)$ values can be seen in Figure~\ref{fig28d}. The four corners of the spherical region having unusual $\hat{H}(t)$ values have HEALPix values 23404309, 23391936, 23564929 and 24158424. Their galactic coordinates were found as $(85.91, -1.66)$, $(76.82, -1.66)$, $(76.82, 4.05)$ and $(85.91, 4.05)$ respectively.

Table~\ref{table26} shows the summary of CMB intensities at these one- and two-dimensional equatorial regions. The two-dimensional region around the unusual values was extracted as a rectangular spherical region from the circular CMB window using the previously identified galactic coordinates to split them as the unusual and the remaining regions. It is clear that the range of temperature intensities is wider in the one- and two-dimensional regions around the unusual values than in the regions excluding them. Further, the variances of intensities in the anomalous regions are larger than in the remaining regions. Moreover, Table~\ref{table26} confirms that the mean $\hat{H}(t)$ values in the anomalous regions are lower than in the remaining regions.

\begin{table}[htb!]
\setlength{\tabcolsep}{4pt}
\centering
\footnotesize
\tabcolsep=0.05cm
\begin{tabular}{@{}*{8}{c}@{}}
\hline
\textbf{\thead{Inspection\\ Window}} & \textbf{\thead{[$I_{\min}$,$I_{\max}$] \\ (in $10^{-3}$)}} & \textbf{\thead{$I_{\max}-I_{\min}$ \\ (in $10^{-3}$)}} & \textbf{\thead{Mean $I$ \\ (in $10^{-5}$)}}  & \textbf{\thead{Variance $I$ \\ (in $10^{-8}$)}} & \textbf{[${\hat{H}(t)}_{\min}$, ${\hat{H}(t)}_{\max}$]} & \textbf{${\hat{H}(t)}_{\min}$-${\hat{H}(t)}_{\max}$} & \textbf{\thead{Mean $\hat{H}(t)$}}\\
\hline
\textbf{\thead{\footnotesize One-dimensional \\ region excluding \\ the region of \\ unusual values}} & {[-0.3688,0.7578]} & {1.1266} & {1.4846} & {1.5654} & {[0.3351, 0.4411]} & {0.1060} & {0.4168} \\ \hline
\textbf{\thead{\footnotesize One-dimensional \\ region around \\ unusual values}} &  {[-0.8865,1.2851]} & {2.1716} & {-9.2156} & {4.9138} & {[0.3336, 0.3496]} & {0.0160} & {0.3384} \\ \hline
\textbf{\thead{\footnotesize Two-dimensional \\ region excluding \\ the region of \\ unusual values}} &  {[-0.3935,0.2721]} & {0.6656} & {-3.6390} & {1.1433} & {[0.4097, 0.7448]} & {0.3351} & {0.5489} \\ \hline
\textbf{\thead{\footnotesize Two-dimensional \\ region around \\ unusual values}} &  {[-0.7310,0.2751]} & {1.0061} & {-6.3779} & {3.6978} & {[0.3015, 0.5975]} & {0.2960} & {0.4398} \\
\hline
\end{tabular}
\caption{Analysis of CMB intensities near the equatorial region}\label{table26}
\end{table}

\begin{figure}[!htb]
    \centering  
    \subfloat[The anomalous sky window]{\label{fig29a}
    \includegraphics[trim={0cm 0.8cm 0cm 1cm},clip,width=0.36\textwidth, height=0.25\textheight]{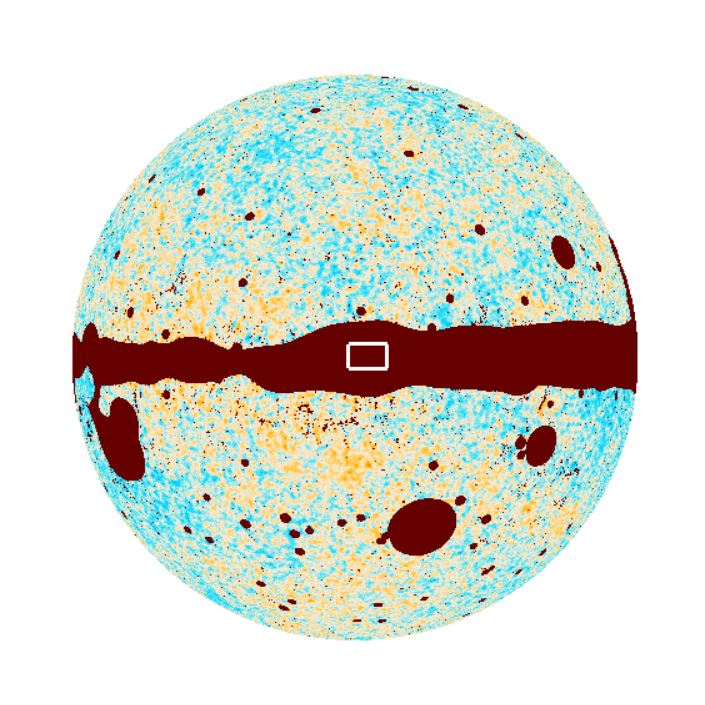}}            
    \centering\hspace{0.4cm}
    \subfloat[The enlarged anomalous sky window]{\label{fig29b}
    \includegraphics[trim={0cm 0cm 0cm 1cm},clip,width=0.30\textwidth, height=0.20\textheight]{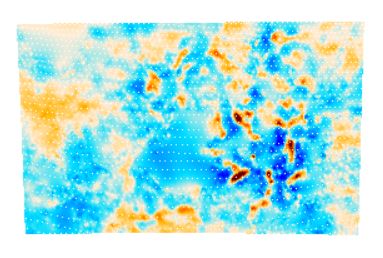}}
    \caption{SMICA 2015 map with TMASK and the region of anomalies}
    \label{fig29}
\end{figure}

Figure~\ref{fig29} shows the Planck 2015 map with blocked non-reliable CMB values. The region where TMASK applied by the SMICA reconstruction technique is removed in Figure~\ref{fig29}. The TMASK of the CMB intensities utilized by the SMICA method determines the region where the inpainted CMB intensities in the galactic plane are considered to be reliable. The rectangular window shows a possible region of anomalies detected by the developed multifractional methodology.

Now we apply this approach and investigate $\hat{H}(t)$ for all $t \in s_2(1)$. First, the one-dimensional methodology was used. $\hat{H}(t)$ was estimated using the CMB intensities on rims, similar to the analysis in Figure~\ref{fig28a} and~\ref{fig28b}. The moving windows with~4096 consecutive pixels, which is approximately a half of a full ring, were used to obtain values of $\hat{H}(t)$. To clearly show local behaviours, after several trials, sets $v_N(t)$ with~61 HEALPix points, i.e. with the radius equals~30 pixels, were selected. The obtained results are shown in Figure~\ref{fig210a}. To compare them with the AC discrepancy approach in~\cite{Hamann:2021}, Figure~\ref{fig210b} shows the corresponding map obtained by applying the direction-dependent probe. The code from~\cite{CMBProbe:2021} was used to compute values of AC discrepancies for SMICA~2015 CMB intensities. The first map highlights $\hat{H}(t)$ values below the~$5^{th}$ percentile. AC discrepancy values above the~${95}^{th}$ percentile were used for the second map. The both approaches detected the region of anomalies in Figure~\ref{fig29}. However, from locations of other discrepancy values, it is clear that these approaches detect different CMB anomalies.

\begin{figure}[!htb]
    \centering  
    \subfloat[H\"older exponent approach]{\label{fig210a}
    \includegraphics[trim={0cm 0.8cm 0cm 0.5cm},clip,width=0.36\textwidth, height=0.25\textheight]{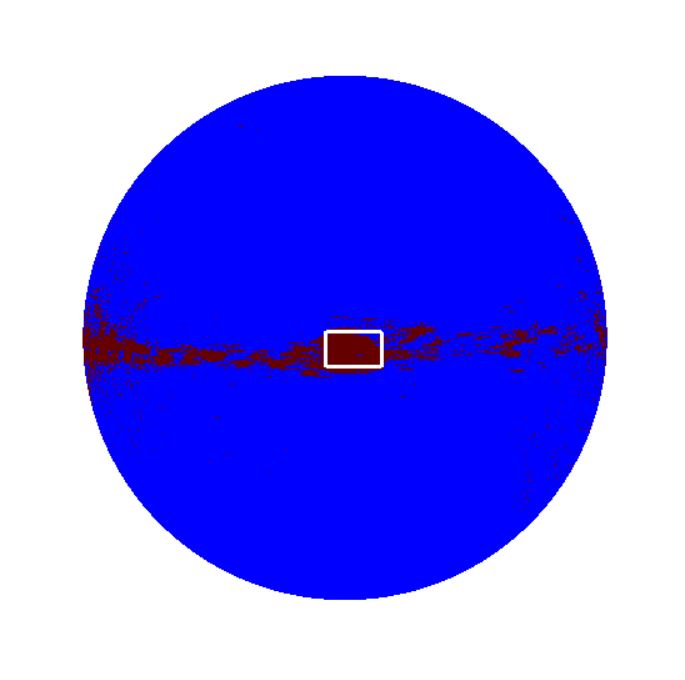}}            
    \centering\hspace{0.4cm}
    \subfloat[AC discrepancy approach]{\label{fig210b}
    \includegraphics[trim={0cm 0.8cm 0cm 0.5cm},clip,width=0.36\textwidth, height=0.25\textheight]{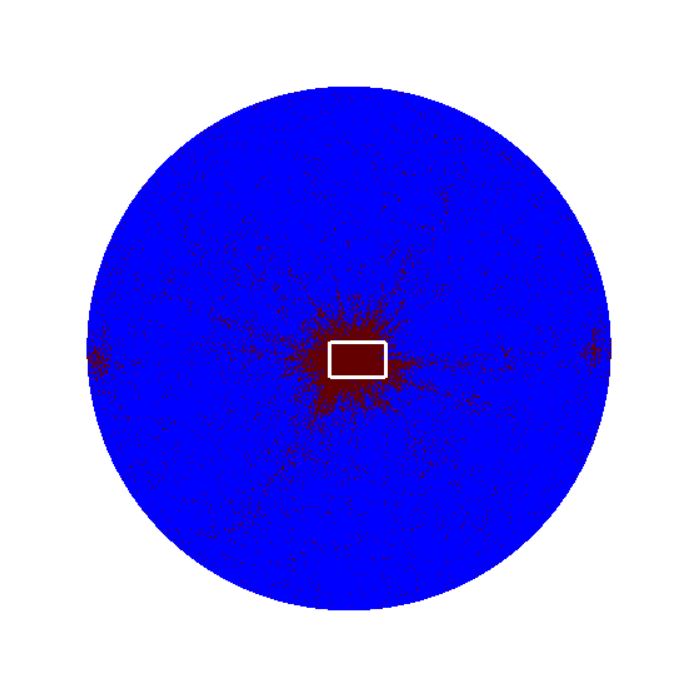}}
    \caption{Discrepancy maps for CMB intensities from SMICA 2015}
    \label{fig210}
\end{figure}

Very sharp changes in $\hat{H}(t)$ values in Figure~\ref{fig28b} motivated the second method to detect anomalies, which is based on increments of $\hat{H}(t)$ values. Figure~\ref{fig28b} demonstrated substantial changes of $\hat{H}(t)$ for nearby $t$ locations. These changes are permanent as $\hat{H}(t)$ exhibits stable behaviour after a rapid \enquote{jump}. Such changes are different from noise or outliers, when values in random distinct locations lay at an abnormal distance from other values in their surrounding points.

To detect such rapid changes, we used the statistics ${\hat{H}}_{\Delta}(t) = \min_{t_1 \in {\Delta}{(t)}}|{\hat{H}}(t)-{\hat{H}}(t_1)|.$
where $t$ and $t_1$ are indices of ring-ordered pixels and the set $\Delta{(t)}$ = $\{t+10,...,t+20\}$. The delay of~10 was selected to detect jumps that occur over short distances. The minimum over the set of consecutive points $\Delta{(t)}$ was used to eliminate outliers or noise that can result in distinct large differences $|{\hat{H}}(t)-{\hat{H}}(t_1)|.$

\begin{figure}[!htb]
    \centering  
    \subfloat[${\hat{H}}_{\Delta}$ discrepancy map]{\label{fig211a}
    \includegraphics[trim={0cm 0.8cm 0cm 0.8cm},clip,width=0.36\textwidth, height=0.25\textheight]{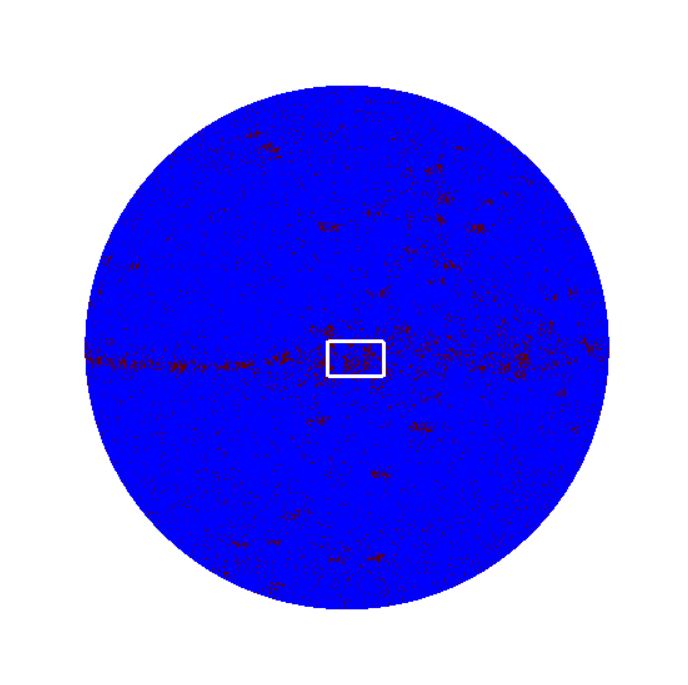}}        
    \centering\hspace{0.4cm}
    \subfloat[${\hat{H}}_{\Delta}$ discrepancies over TMASK]{\label{fig211b}
    \includegraphics[trim={0cm 0.8cm 0cm 0.8cm},clip,width=0.36\textwidth, height=0.25\textheight]{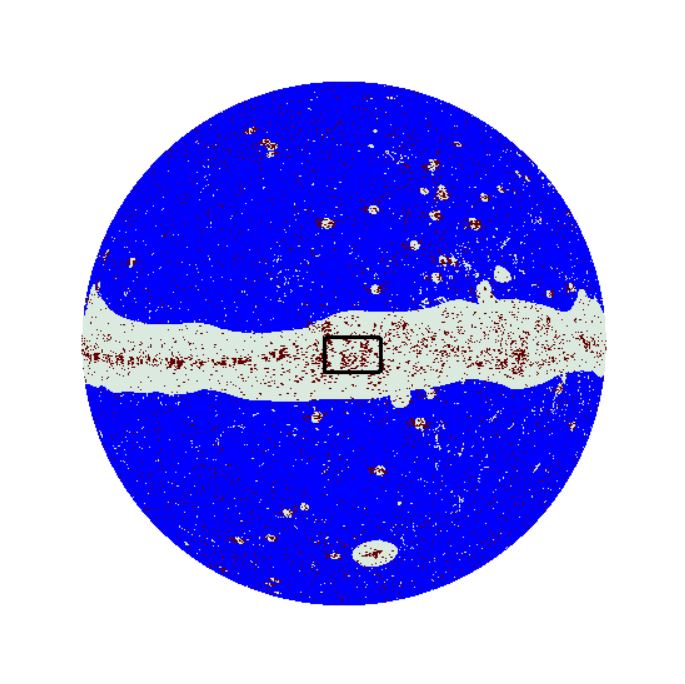}}
    \caption{${\hat{H}}_{\Delta}$ discrepancy maps for CMB intensities from SMICA 2015}
    \label{fig211}
\end{figure}

Figure~\ref{fig211a} shows the computed ${\hat{H}}_{\Delta}(t)$ values for SMICA~2015 CMB intensities. ${\hat{H}}_{\Delta}(t)$ values above the~${95}^{th}$ percentile are plotted.

In Figure~\ref{fig211b}, $5\%$ of largest ${\hat{H}}_{\Delta}(t)$ values are shown on the TMASK map. It can be seen that in most cases, clusters of largest ${\hat{H}}_{\Delta}(t)$ values are within the TMASK. It seems that ${\hat{H}}_{\Delta}$ statistics rather accurately detected many regions with unreliable CMB values. Analysis of other CMB maps gave similar results. 

Summarising, the implemented methodology to investigate multifractional presence within the CMB data, could also serve as a mechanism to detect regions of anomalies in CMB maps.

\section{Conclusion}
\label{S2:7}

This paper examined multifractional spherical random fields and their applications to analysis of cosmological data from the Planck mission. It estimated pointwise H\"older exponent values for the actual CMB data and checked for the presence of multifractionality. The estimators of pointwise H\"older exponents for one- and two-dimensional regions were obtained by using the ring and nested orderings of the HEALPix visualization structure. The carried out analysis conveyed some multifractionality in the CMB data since the computed pointwise  H\"older exponent values do change from  place to place in the CMB sky sphere. The proposed approach was also applied to introduce statistics that were used for detecting regions with anomalies in CMB data. The developed methodology can be used for other spherical data.

Some numerical approaches that were used to speed up computations for big CMB data sets will be reported in detail in future publications.
In future studies, it would be also interesting to:
\begin{itemize}[-]
    \item Develop the distribution theory for the estimators of $H(t)$;
    \item Develop  hypothesis tests of equality of the local H\"older exponents taking into account the dependence structure of  random fields; 
    \item Investigate reliability and accuracy of various estimators of the H\"older exponent for CMB data;
    \item Study rates of convergence in Theorem~\ref{theo2.4.1};
    \item Investigate changes of the H\"older exponents depending on evolutions of random fields driven by SPDEs on the sphere, see~\cite{Anh:2018, Broadbridge:2019, Broadbridge:2020, RESTREPO:2021};
    \item Study directional changes of the H\"older exponent by extending the obtained results for the conventional ring ordering to rings with arbitrary orientations;
    \item Apply the developed methodology to other spherical data, in particular, to new high-resolution CMB data from future CMB-S4 surveys~\cite{CMBS4:2019};
    \item Explore relations between the locations of the detected CMB anomalies and other cosmic objects.
\end{itemize}

\section*{Declaration of competing interest}

The authors declare that they have no known competing interests for the results reported in this paper. 

\section*{Author statement}

All the authors equally contributed to the paper.

\section*{Acknowledgments}

This research was partially supported under the Australian Research Council's Discovery Projects funding scheme (project number  DP160101366). We would like to thank Professor Antoine Ayache for attracting our attention to and discussing mutifractional models for random fields and Professor Ian~Sloan for various discussions about mathematical modelling of CMB data.

This research includes computations using the Linux computational cluster Gadi of the National Computational Infrastructure (NCI), which is supported by the Australian Government and La Trobe University. We are also grateful for the use of data of the
Planck/ESA mission from the Planck Legacy Archive.
 
\bibliographystyle{elsarticle-num}
\bibliography{Bibliography}

\end{document}